\documentclass[12pt,journal,onecolumn,doublespacing]{IEEEtran}
\usepackage{url}
\usepackage{tocloft}
\usepackage{setspace}
\usepackage{amsmath}
\usepackage{multicol}
\usepackage{amssymb}
\usepackage[utf8]{inputenc}
\usepackage[T1]{fontenc}
\usepackage[font={small}]{caption}
\usepackage{cite}
\usepackage{enumerate}
\usepackage[square, comma, numbers, sort&compress]{natbib}
\usepackage{subfig}
\usepackage{bm}
\usepackage{multirow}
\usepackage{tabularx}
\usepackage{subfig}
\usepackage{upgreek}

\usepackage{url}
\usepackage{tocloft}
\usepackage{setspace}
\usepackage{amsmath}
\usepackage{multicol}
\usepackage{amssymb}
\usepackage[utf8]{inputenc}
\usepackage[T1]{fontenc}
\usepackage[font={small}]{caption}
\usepackage{cite}
\usepackage{enumerate}
\usepackage[square, comma, numbers, sort&compress]{natbib}
\usepackage{subfig}
\usepackage{bm}
\usepackage{multirow}
\usepackage{algorithm}
\usepackage{algorithmic}
\usepackage{float}% http://ctan.org/pkg/float

\newtheorem{proposition}{Proposition}

\makeatother

\providecommand{\theoremname}{Theorem}

\makeatother

\providecommand{\lemmaname}{Lemma}

\makeatother

\providecommand{\propositionname}{Proposition}

\makeatother

\providecommand{\corname}{Corollary}

\makeatother

\providecommand{\remname}{Remark}

\ifCLASSINFOpdf
\else
\fi
\ifCLASSINFOpdf
   \usepackage[pdftex]{graphicx}
   \usepackage{epstopdf}
\else
\fi

\begin{document}
	
	%\pagenumbering{gobble}
	
	\title{Toward Autonomous Reconfigurable Intelligent Surfaces Through Wireless Energy Harvesting}

	%\author{\IEEEauthorblockN{Konstantinos Ntontin\IEEEauthorrefmark{1}, Marco Di Renzo\IEEEauthorrefmark{2}, and Fotis Lazarakis\IEEEauthorrefmark{1}
%}\\\IEEEauthorblockA{\IEEEauthorrefmark{1} Institute of Informatics and Telecommunications, National Centre for Scientific Research "Demokritos", \\15310, Agia Paraskevi, Greece, E-mail:\{konstantinos.ntontin, flaz\}@iit.demokritos.gr \\\IEEEauthorrefmark{2} Laboratoire des Signaux et Syst\`{e}mes, CNRS, CentraleSup\'{e}lec, Univ Paris Sud, Universit\'{e} Paris-Saclay, 3 rue Joliot Curie, Plateau du Moulon, 91192, Gif-sur-Yvette, France, E-mail: marco.direnzo@l2s.centralesupelec.fr}}

\author{Konstantinos Ntontin, \IEEEmembership{Member, IEEE}, {Alexandros--Apostolos A. Boulogeorgos}, \IEEEmembership{Senior Member, IEEE}, Emil Bj\"{o}rnson, \IEEEmembership{Senior Member, IEEE},  Dimitrios Selimis, \IEEEmembership{Graduate Student Member, IEEE}, Wallace Alves Martins, \IEEEmembership{Senior Member, IEEE}, Steven Kisseleff, \IEEEmembership{Member, IEEE}, Sergi Abadal, \IEEEmembership{Member, IEEE}, Eduard Alarc\'{o}n, \IEEEmembership{Senior Member, IEEE}, Anastasios Papazafeiropoulos, \IEEEmembership{Senior Member, IEEE},  Fotis Lazarakis,   
	Angeliki Alexiou, \IEEEmembership{Member, IEEE}, and Symeon Chatzinotas, \IEEEmembership{Senior Member, IEEE}
	
		\thanks{K. Ntontin, W. A. Martins, S. Kisseleff, and S. Chatzinotas are with the Interdisciplinary Centre for Security, Reliability and Trust (SnT) – University of
		Luxembourg, L-1855 Luxembourg. E-mail: $\left\{\text{konstantinos.ntontin, wallace.alvesmartins, steven.kisseleff, symeon.chatzinotas}\right\}$@uni.lu.}
	
	\thanks{A.-A. A. Boulogeorgos and A. Alexiou  are with the Department of Digital Systems,
		University of Piraeus
		Piraeus 18534 Greece. E-mail:  al.boulogeorgos@ieee.org,  alexiou@unipi.gr.}
		
		\thanks{E. Bj\"{o}rnson is with the Department of Electrical Engineering, Link\"{o}ping University, 581 83 Link\"{o}ping, Sweden, and with the Department of Computer Science, KTH, 164 40 Kista, Sweden. Email: emilbjo@kth.se.}
	
	\thanks{D. Selimis and F. Lazarakis are with the Wireless
		Communications Laboratory of the Institute of Informatics and Telecommunications, National Centre for Scientific Research–``Demokritos,'' Athens, Greece. E-mails: $\left\{\text{dselimis, flaz}\right\}$@iit.demokritos.gr.}
	
	\thanks{S. Abadal and E. Alarc\'{o}n are with the NaNoNetworking Center in Catalunya (N3Cat), Universitat Politècnica de Catalunya, 08034 Barcelona, Spain. Emails: abadal@ac.upc.edu, eduard.alarcon@upc.edu.}
	
	\thanks{A. Papazafeiropoulos is with the Communications and Intelligent Systems Research Group, University of Hertfordshire, Hatﬁeld AL10 9AB, U.K., and also with the SnT, University of Luxembourg, 4365 Luxembourg City, Luxembourg. E-mail: tapapazaf@gmail.com.}
\thanks{This work was supported by the European Commission's Horizon 2020 research and innovation programme (ARIADNE) under grant agreement No. 871464 and the Luxembourg National Research Fund (FNR) under the CORE project RISOTTI.} }

	\maketitle
	
	\begin{abstract}
	    In this work, we examine the potential of autonomous operation of a reconfigurable intelligent surface (RIS) using wireless energy harvesting from information signals. To this end, we first identify the main RIS power-consuming components and introduce a suitable power-consumption model. Subsequently, we introduce a novel RIS power-splitting architecture that enables simultaneous energy harvesting and beamsteering. Specifically, a subset of the RIS unit cells (UCs) is used for beamsteering while the remaining ones absorb energy. For the subset allocation, we propose policies obtained as solutions to two optimization problems. The first problem aims at maximizing the signal-to-noise ratio (SNR) at the receiver without violating the RIS's energy harvesting demands. Additionally, the objective of the second problem is to maximize the RIS harvested power, while ensuring an acceptable SNR at the receiver. We prove that under particular propagation conditions, some of the proposed policies deliver the optimal solution of the two problems. Furthermore, we report numerical results that reveal the efficiency of the policies with respect to the optimal and very high-complexity brute-force design approach. Finally, through a case study of user tracking, we showcase that the RIS power-consumption demands can be secured by harvesting energy from information signals.
	\end{abstract}

		% no keywords

\section{Introduction}
\label{Introduction}

The data traffic demands increase steadily and exponentially \cite{5G_Vision_Cisco}. To prevent a possible capacity crunch, a key proposed solution is traffic offloading to higher frequency bands \cite{Andrews_what_will_5G_Be}. More specifically, a large fraction of the traffic in future access networks is expected to utilize the millimeter-wave (mmWave) spectrum due to the larger available bandwidth \cite{Rappaport_Wideband_MmWave_Measurements}. However, this advantage comes with a cost: mmWave bands are more susceptible to fixed and moving blockages in comparison to their sub-6-GHz counterparts due to the high penetration loss that may even reach 40 dB for certain materials, such as tinted glass \cite{Reflection_penetration_loss_28_GHz}.

A possible solution to the blockage problem that has extensively been discussed in the literature is the creation of alternative signal paths through relays or passive reflectors. Relays can undoubtedly enhance the coverage by enabling the transmitted signal to be rerouted through them when the direct link is blocked. However, since relays require power amplifiers and possibly baseband processing, their power consumption will be comparable to that of mmWave small-cell base stations.
 This fact makes their massive deployment in mmWave networks questionable \cite{Khawaja_mmWave_Passive_Reflectors}. On the other hand, passive reflectors can enhance the coverage in mmWave networks, while requiring no power supply \cite{Khawaja_mmWave_Passive_Reflectors}, \cite{THz_Passive_Reflectors}. However, the shape of the reflected signal is determined when they are deployed (e.g., based on Snell's law for a planar homogeneous reflector), and thus it cannot be adapted based on user mobility or changes in the propagation environment.

To circumvent the aforementioned limitations of relays and passive reflectors, reconfigurable intelligent surfaces (RISs) have been identified as possible countermeasures  \cite{Marco_Di_Renzo_Survey_RISs}, \cite{Emil_Myths_RISs}. RISs are artificial structures that usually consist of a dielectric substrate, which embeds conductive elements named unit cells (UCs), of sub-wavelength size and spacing. Typical UCs are composed of dipoles, patches, and string resonators. By properly tuning their impedance through semiconductor components, such as positive-intrinsic-negative (PIN) diodes,  field-effect transistors (FETs), and radio-frequency microelectromechanical systems (RF-MEMS), the amplitude and phase response to an impinging electromagnetic wave can be altered \cite{Tsilipakos2020a}. In particular, through such adjustment, different functionalities can be performed: beam steering toward a desired angular direction or toward a desired point, beam splitting, and absorption. In terms of power consumption, an amount of power is consumed for their reconfigurability, which arguably is significantly lower compared to relays' energy consumption, at least if the duty cycle of reconfiguration is low \cite{Emil_Myths_RISs}. This has led to the characterization of RISs as nearly passive structures \cite{Passive_Beamforming}, \cite{Xiaojun_Yuan_2021}.

Inspired by the aforementioned properties, a notable amount of research efforts has been devoted to characterizing the RIS functionalities, providing end-to-end channel models and channel estimation schemes, optimizing the resource allocation and discussing access, localization and wireless power transfer approaches in RIS-empowered systems, as well as verifying them through experimental measurements; see
 \cite{Marco_Di_Renzo_Survey_RISs} for a recent survey. In addition, their comparison with the most competitive technology that also allows beamsteering to an arbitrary direction, which is active relaying, has been examined in several works, which showcase that sufficiently large RISs can outperform their relaying-based approaches \cite{Comparison_RIS_Relaying_Boulogeorgos, Renzo2020, RIS_Scaling_Laws_ERmil_Bjornson, Huang_Reconf_Intell_Surfaces, Bjornson_Relaying}. In terms of wireless power transfer, there have been several works considering the deployment of RISs for powering energy-constrained users by taking into account the fact that a large RIS can collect a sufficient amount of energy arriving from the transmitter and direct it to the particular users \cite{Zheng_2020_WPT, Pan_2020_SWIPT, Wu_2020_SWIPT, Shi_2020_WPT, Yang_2021_WPT_RISs, Bouanani_2021_WPT_RISs, Chu_2021_WPT_RISs, Tong_2021_WPT_RISs, Psomas_2021_WPT_RISs, Zargari_2021_WPT_RISs, Yu_2021_WPT_RISs}. 
 
 Building on the above considerations, in this paper we examine the case for \emph{autonomous RIS operation}, which we define as a mode where the RIS does not require a dedicated power supply.
 To the best of our knowledge, as far as the autonomous RIS operation is concerned, only \cite{Wirelessly_Powered_RISs_TVT} and \cite{Chu_2021_WPT_RISs_2} consider the possibility of harvesting energy from the impinging wireless signals to satisfy the power consumption needs of the RISs. In particular, both works consider a time-splitting architecture in which specific time slots are allocated for only wireless energy harvesting from the RIS, while the remaining ones are allocated for data transmission assisted by the RIS, which uses the previously harvested energy to reconfigure itself during beamsteering operations. 

\textbf{Motivation, novelty, and contribution}:  Autonomous RIS operation has been suggested as a possibility due to its expected low power requirements. However, to the best of our knowledge, no work up until now has quantified the particular requirements under practical scenarios of user mobility that require the periodical reconfiguration of such a surface. To quantify the particular amount, we first need to identify the main RIS power-consuming electronic components. This is crucial in order to determine accurate power consumption values of the RIS electronics, based on realistic use cases, and to verify if and when an autonomous RIS operation is plausible through wireless energy harvesting. Furthermore, the time-splitting architectures proposed in \cite{Wirelessly_Powered_RISs_TVT} and \cite{Chu_2021_WPT_RISs_2} would require a protocol redesign of the corresponding networks so that certain time slots are allocated only for the wireless energy harvesting process at the RISs. In addition, time-splitting architectures achieve worse resource utilization efficiency compared to their power-splitting counterparts, since in the latter case the transmission of information is not periodically interrupted by power transmissions. Based on the aforementioned issues, our contribution in this work can be summarized as follows:

\begin{itemize}
	\item We identify the main RIS power-consuming modules and, based on them, we introduce the corresponding power-consumption model. 
	
	\item To overcome the aforementioned difficulty of protocol redesign and reduced resource utilization efficiency in the case of time-splitting harvesting architectures, we introduce a novel power splitting-based architecture in which a part of the total electromagnetic power impinging on the RIS is harvested and, simultaneously, the rest is steered toward the end-user. This is achieved by dynamically allocating, based on the channel conditions, a subset of the UCs for energy harvesting while the rest for beamsteering.
	
	\item For the simultaneous RIS energy harvesting and beamsteering system, we derive closed-form expressions for the end-to-end signal-to-noise ratio (SNR) and direct current (DC) harvested power.
	
	\item Relying on the aforementioned expressions, we formulate two optimization problems that output sets of UCs that need to be used for harvesting and beamsteering. The first optimization problem aims at maximizing the SNR at the end-user while concurrently satisfying the RIS's power-consumption demands. The second optimization problem targets the maximization of the RIS harvested power, while ensuring an acceptable SNR at the end-user. For both problems, we provide closed-form sub-optimal policies for their solution and prove that, under particular propagation conditions, some of the policies deliver the optimal allocation.
	
	\item To validate the efficiency of the proposed solutions, we perform extensive Monte-Carlo simulations and compare their performance against the brute-force approach. In addition, we substantiate the claim that autonomous RIS operation can be feasible through a case study of user tracking that delivers the required RIS power consumption under practical assumptions.
	
\end{itemize}

The rest of this work is structured as follows.\footnote{This work is an extension of \cite{Ntontin_Autonomous_RIS_operation_Submission_Globecom_2021}. In particular, in \cite{Ntontin_Autonomous_RIS_operation_Submission_Globecom_2021} only the first problem of the SNR maximization is considered. In addition, instead of an arbitrary fading model that we consider in this work, which makes the analysis more flexible, the special-case of free-space propagation is considered for both the transmitter-RIS and RIS-receiver links.} In Section~\ref{System_Model}, the system and channel models are presented together with the identification of the main RIS modules that consume power and the resulting power-consumption model. In Section~\ref{Harvested_Power}, the instantaneous DC harvested power and the end-to-end SNR are computed and the two optimization problems of interest are formulated. In Section~\ref{problem_solution}, closed-form sub-optimal policies for the solution of the problems are provided followed by a proof that under certain propagation conditions some of the policies deliver the optimal allocation. In addition, in Section~\ref{Numerical_results}, a case study of RIS-assisted user tracking is presented that verifies the RIS autonomous operation potential under certain conditions. Finally, Section~\ref{Conclusions} concludes this work.

\section{System, Channel, and RIS Power-Consumption Models}

\label{System_Model}

In this section, we first present the system and channel models under consideration. Subsequently, we identify the power-consuming RIS modules and, based on them, we introduce the corresponding RIS power-consumption model.

\subsection{System model}

\label{System Model}

\subsubsection{Scenario}

\begin{figure}
	%\label{Communication_through_an_RIS}
	\centering
	{\includegraphics[width=\columnwidth]{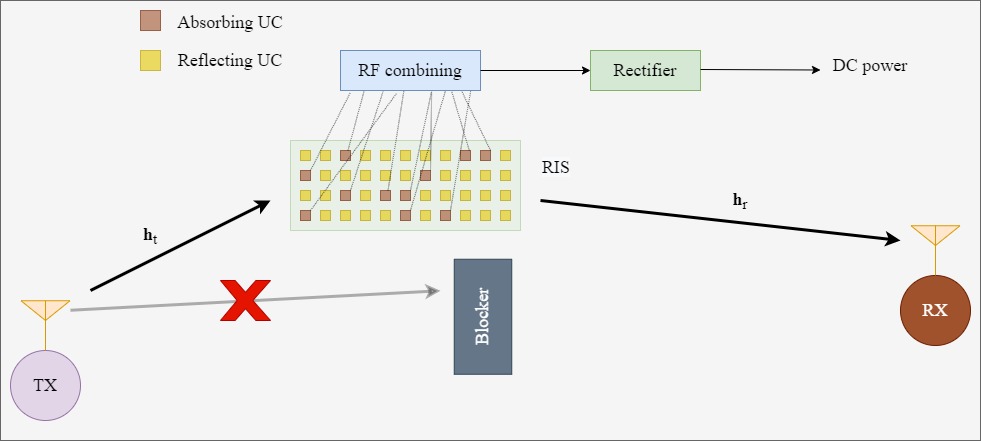}}
	\caption{Scenario and energy-harvesting architecture.}
	%\hrulefill
	\label{Fig:Communication_through_an_RIS}	
\end{figure}

As illustrated in Fig.~\ref{Fig:Communication_through_an_RIS}, we consider a scenario in which a transmitter (TX) communicates with a receiver (RX) through a planar RIS located in the far-field of both the TX and RX. 
%The incidence and departure angles of the electromagnetic wave with respect to the center of the illuminated area, which is assumed to coincide with the RIS center, are respectively denoted by $\theta_{\rm{inc}}$ and $\theta_{\rm{dep}}$.
The TX-RIS link, of distance $d_t$\,m, and RIS-RX link, of distance $d_r$\,m, constitute an alternative path to the direct TX-RX link that is assumed to be blocked. 
%In addition, we assume that the TX and RX antennas are pointing toward the center of the illuminated RIS region with gains $G_t$ and $G_r$, respectively. 
The RIS is a rectangular uniform planar array consisting of $M_s=M_x\times M_y$ UCs of size $d_x \times d_y$. $M_x$ ($M_y$) and $d_x$ ($d_y$) denote the number of UCs and their length in the x-axis (y-axis), respectively.

%Moreover, we consider that each UC is an electrically-small low-gain element with gain pattern, which is supported by measurements \cite{Tang_measurements_2021}, that can be expressed as
%\begin{align}
%	\label{RIS_element_gain}
%	G_{s}\left(\theta\right)= 4{\cos\left(\theta\right)}, \text{ with } 0\le\theta<\pi/2.
%\end{align}

\subsubsection{Energy-harvesting architecture}

Among the $M_s$ UCs, through the impedance-adjusting semiconductor components attached to them, $M_h$ UCs are configured to act as perfect absorbers of the impinging electromagnetic energy, which is used for supplying the RIS modules. Furthermore, we consider a corporate-feed network approach in which the radio frequency (RF) harvested power of the harvesting UCs is combined and driven to a single rectifier circuit through the particular network \cite{ElBadawe2017}, \cite{ElBadawe2018}. The benefit of such an approach, compared to a solution in which a rectifier is attached to each UC and, subsequently, the DC harvested powers are combined, is the maximization of the input RF power at the RF circuit together with the minimization of the circuitry required for rectification. The former could be an essential requirement since the harvested power of a single UC might be low for turning on the diode of the rectifying circuit in an approach where such a circuit is attached to each UC. On the other hand, combining the harvested RF powers through the network could result in lower radiation to alternative current radiation efficiency compared to the one rectifier-per-UC approach \cite{Amer_Metasurface_Energy_Harvesting}. 

The remaining $M_r = M_s - M_h$ UCs, through again the impedance-adjusting semiconductor components, are configured to act as perfect reflectors of the impinging energy, beamforming it toward the RX. We consider that $M_h$ and $M_r$ are not fixed, but can dynamically change according to the channel conditions. To achieve this, it is assumed that the UCs that are selected for harvesting during a reconfiguration interval are connected to the corporate-feed network through RF switches integrated into the RIS structure.

\subsection{Channel model}

We assume a flat-fading channel model where the complex envelope channel vectors of the TX-RIS and RIS-RX links  denoted by $\textbf{h}_t$ and $\textbf{h}_r$, are given by
\begin{equation}
	\label{channel_model1}
	{\bf{h}}_{\rm{t}}=[h_{t_1}\quad h_{t_2}\quad \cdots \quad h_{t_{M_s}}]^{T}, \quad {\bf{h}}_{\rm{r}}=[h_{r_1}\quad h_{r_2}\quad \cdots \quad h_{r_{M_s}}]^{T},
\end{equation} where the UCs can be indexed arbitrarily. These vectors describe the joint effect of antenna gains, geometric pathloss, and multipath fading (i.e., the combination of small-scale and large-scale fading).

The channel vectors ${\bf{h}}_{\rm{t}}$ and ${\bf{h}}_{\rm{r}}$ are time-varying due to mobility and multipath fading, which calls for a time-varying RIS configuration. We consider a block-fading model where the channel vectors are fixed within fixed-size time intervals, but change abruptly between intervals. Hence, the RIS must be reconfigured in each interval. The duration of these intervals depends on the time-variation of the channel. The optimization problems and policies developed in this article operate on a per-interval basis, thus they can be applied irrespective of how the channel realizations are generated. For example, a stationary fading distribution such as Rayleigh, Nakagami-m, Weibull, or Rice could be used, but also a deterministic ray-traced or measured channel evolution could be considered.

To focus on the power consumption modeling, we assume that the TX has perfect channel state information knowledge of both the TX-RIS and RIS-RX links, which leverages for the reconfiguration commands that periodically transmits to the RIS by wireless pilot signals. In addition, in order to maintain a low-complexity RIS structure suitable for autonomous operation, we assume that neither channel estimation can be performed in the RIS, due to lack of the required circuitry, nor pilot signals can be sent by the RIS to the TX and RX for channel estimation \cite{Two_timescale_channel_estimation}. Although this is a challenging scenario for separate estimation of the TX-RIS and RIS-RX channels instead of the cascaded TX-RIS-RX one, methods have been reported that leverage the quasi-static nature of the TX-RIS link, in scenarios where both the TX and RIS are fixed, to achieve such estimation \cite{Two_timescale_channel_estimation}. Moreover, we  note that although no channel estimation performed by the RIS is assumed, the RIS would still consume an amount of power for channel estimation either at the TX or RX. This is due to the fact that during the channel estimation training periods some of UCs would need to be reconfigured based on phase patterns known at the node that performs the estimation (either the TX or RX) \cite{Two_timescale_channel_estimation}. We assume that such RIS power consumption demands can be covered by a harvested power surplus during the information transmission periods. We further note the need for frequent updates of channel state information is expected to be alleviated in scenarios of either low mobility or dominant LoS links. This, in turn, could considerably reduce RIS power consumption demands for channel estimation.

\subsection{Power-consumption model}

\label{Power_characteristics_semiconductor_components}

To present a suitable RIS power consumption model, we first need to identify the main RIS power-consuming modules, namely: i) impedance-adjusting semiconductor components, ii) control network, and iii) rectifier. They are described below.

\subsubsection{Impedance-adjusting semiconductor components}

 The power consumption of these components is characterized by static and dynamic factors. The static factor corresponds to their uninterrupted power consumption due to leakage currents originating from the bias voltages when they operate in steady state. Usually, this factor is negligible for FETs and RF MEMS \cite{Paradigm_Phase_Shift}. On the other hand, the dynamic factor constitutes a non-negligible factor related to the charging and discharging of internal capacitors during bias voltage level changes. This is needed for UC phase and amplitude response adjustment. It appears only when the semiconductor components change state.

\subsubsection{Control network}

\label{Characteristics_of_the_control_network}

As described in \cite{Emil_Myths_RISs}, \cite{Abadal_Programmable_Metamaterials}, the RIS needs to receive external commands regarding the new configuration states for the UCs. This can be achieved by one of the following basic approaches: i) detached microcontroller architecture and ii) integrated architecture. The first one is the common architecture that has been used for several years in which the control network is realized by a microcontroller, such as a field-programmable gate array (FPGA). Such architectures are usually bulky and subject to significant power consumption, thus reducing the potential for RIS autonomous operation \cite{Abadal_Programmable_Metamaterials}. In contrast to the FPGA-based architectures, the control network in the integrated architecture is realized by a network of communicating chips (usually one per UC) that read the UC state and adjust the bias voltages of the impedance-adjusting semiconductor elements. These circuits receive, interpret, and apply the commands  and exhibit their own static and dynamic power consumption due to leakage and transistor switching, respectively \cite{Tasolamprou_exploration_Intercell}. In addition, they are likely to use asynchronous logic as it does not require a complex and power-hungry clock signal distribution \cite{Asynchronous_Logic_Hypersurface}. Furthermore, it has been suggested that such chips could also operate autonomously through energy harvesting, based on nano-networking advancements \cite{Abadal_Programmable_Metamaterials}, \cite{Liaskos_2018_Realizing_Wireless}.

\subsubsection{Rectifier}

\label{Rectifier}

For the RF-to-DC power conversion that is needed to power the electronic modules of the RIS, a rectifier circuit follows the RF combiner. This circuit can be either passive that exhibits negligible power consumption, or active, by incorporating active diodes with lower voltage drop that can increase the conversion efficiency \cite{Perera_SWIPT}. In the latter case, the rectifier exhibits a non-negligible power consumption.

To minimize the RIS consumption, we consider an integrated RIS architecture. Moreover, it is assumed that each UC is connected to one chip that controls and adjusts its impedance through an impedance-adjusting semiconductor component \cite{Tasolamprou_exploration_Intercell}. By assuming that only passive rectifiers are employed on the RIS and denoting its power consumption by $P_{\rm{RIS}}$, it holds that \cite[Eq. (4.5)]{D_5_3_VISORSURF}
\begin{align}
	\label{Total_RIS_power}
	P_{\rm{RIS}}=P_{\rm{static}}^{\rm{tot}}+P_{\rm{dynamic}}^{\rm{tot}},
\end{align} where $P_{\rm{static}}^{\rm{tot}}$ and $P_{\rm{dynamic}}^{\rm{tot}}$ denote the total static and dynamic RIS power consumption that incorporates the consumption arising from both the control chips and impedance-adjusting semiconductor components.

Regarding $P_{\rm{static}}^{\rm{tot}}$, under the assumption of one control chip and one impedance-adjusting semiconductor component per UC, it is expected to scale linearly with the number of unit cells. Hence, we have
\begin{align}
	\label{total_static_power_consumption}
	P_{\rm{static}}^{\rm{tot}}=M_sP_{\rm{static}},
\end{align}
where $P_{\rm{static}}$ denotes the accumulated static power consumption of a control chip and an impedance-adjusting semiconductor component.

As far as $P_{\rm{dynamic}}^{\rm{tot}}$ is concerned, not necessarily all UCs need to change their response during a particular reconfiguration period. In particular, through a case study of user tracking, it is revealed in \cite{Saeed2019} that different number of UCs need to be reconfigured for different relative positions of the moving user with respect to the RIS. Hence, $P_{\rm{dynamic}}^{\rm{tot}}$ is given by \cite[Eq. (8.11)]{liaskos2020internet} as
\begin{align}
	\label{total_dynamic_power_consumption}
	P_{\rm{dynamic}}^{\rm{tot}}=M_s\alpha p_rP_{\rm{dynamic}}=M_sP_{d}^{\rm{avg}},
\end{align} where $\alpha \in \left(0,1\right]$ is the probability for a UC to change its state, $p_r$ is the percentage of time that a UC is subject to reconfiguration, and $P_{\rm{dynamic}}$ is the aggregated power consumption during reconfiguration of the control chips and impedance-adjusting semiconductor components. The term $P_{d}^{\rm{avg}}=\alpha p_rP_{\rm{dynamic}}$ can be viewed as an equivalent continuous power consumption per control chip that corresponds to its dynamic operation. By plugging \eqref{total_static_power_consumption} and  \eqref{total_dynamic_power_consumption} into \eqref{Total_RIS_power}, we obtain
\begin{align}
	P_{\rm{RIS}}=M_s\left(P_{\rm{static}}+P_{d}^{\rm{avg}}\right).
\end{align}

\section{DC harvested power, end-to-end SNR, and problem formulation}

\label{Harvested_Power}

The aim of this section is to first present and quantify the autonomous RIS-empowered system design parameters. Subsequently, we use them to formulate the optimization problems of interest. In more detail, Section~\ref{DC_Harvested_Power} returns a closed-form expression for the harvested power, Section~\ref{End_to_end_SNR} reports the end-to-end SNR and, finally, Section~\ref{Problem_formulation} describes the formulation of the two problems. 

\subsection{DC harvested power}

\label{DC_Harvested_Power}

For the DC harvested power, we use the non-linear model from \cite{Boshkovska_Non_Linear_EH_Model}, which has been extensively validated through experimental measurements. According to this model, the output DC power of the rectifier can be evaluated as
\begin{align}
	\label{DC_power}
	P_{\rm{DC}}=\frac{\frac{P_{\rm{max}}}{1+e^{-a\left(P_{\rm{harv}}-b\right)}}-\frac{P_{\rm{max}}}{1+e^{ab}}}{1-\frac{1}{1+e^{ab}}},
\end{align}
where $P_{\rm{harv}}$ is the RF power that is inputted to the rectifier and $P_{\rm{max}}$ is a constant denoting the maximum harvested power in the case that the harvesting circuit at the rectifier is saturated. In addition, $a$ and $b$ are circuit-specific parameters, which are related to the resistance, capacitance, and turn-on voltage of the diode used for rectification \cite{Boshkovska_Non_Linear_EH_Model}.

As far as $P_{\rm{harv}}$ is concerned, let us denote the set of the UCs that are selected for energy harvesting by $\mathcal{A}_h$. 
Since the transmit power is $P_t$, the $i_{\rm{th}}$ UC, $i \in \mathcal{A}_h$, will receive a signal $y_i = \sqrt{P_t} h_{t_i}$, where the thermal noise has been omitted since it cannot be harvested.
Hence, the total harvested power as a function of  $\mathcal{A}_h$ can be expressed as
\begin{align}
	\label{Harvested_power}
	P_{\rm{harv}} = \eta_{\rm{RF}} P_{t} \sum_{i \in \mathcal{A}_h}\left|h_{t_i}\right|^2,
\end{align}
where $\eta_{\rm{RF}} \in (0,1]$ denotes the efficiency of combining the individual harvested powers from each UC through the corporate-feed network. 
We note that according to \eqref{Harvested_power} only the power corresponding to the transmission from the TX is taken into account in the harvesting process. This is due to the assumption that the radiation pattern of the RIS corresponding to the harvesting UCs is aligned with the direction of the TX main lobe (that is directed toward the RIS center) in order to maximize the absorption of the power arriving from the particular direction. For this reason, it is expected that the absorbed power level of ambient signals arriving to the RIS from other directions would be negligible compared to the one arriving from the TX. 

\subsection{End-to-end SNR}

\label{End_to_end_SNR}

Let $\mathcal{A}_r$ denote the set of UCs that are selected for reflection, which is the orthogonal complement of $\mathcal{A}_h$, i.e., $\mathcal{A}_r=\mathcal{A}_h^C$. The end-to-end SNR $\gamma$ can be obtained by following the standard approach as
\cite{Huang_Reconf_Intell_Surfaces,Bjornson_Relaying,RIS_optimal_placement_Ntontin}
\begin{align}
	\label{received_power}
	\gamma=\frac{P_{t}}{\sigma^2} %\left(\frac{\lambda}{4\pi}\right)^4\frac{P_{t}G_{t}G_rG_{s}\left(\theta_{\rm{inc}}\right)G_{s}\left(\theta_{\rm{dep}}\right)}{d_t^2d_r^2}
	\left|\sum_{k \in \mathcal{A}_r}h_{t_k}h_{r_k}e^{j\left(\phi_k+\angle{h_{t_k}}+\angle{h_{r_k}}\right)}\right|^2,
\end{align}
where $\phi_k$ is the induced phase response from the $k_{\rm{th}}$ UC, $\angle$ denotes the angle of the corresponding complex number, and $\sigma^2$ is the variance of the thermal noise at the RX. From \eqref{received_power} it is evident that the SNR is maximized by setting
\begin{align}
	\label{optimum_phases}
	\phi_k=-\angle{h_{t_k}}-\angle{h_{r_k}},\; \text{for} \quad k \in \mathcal{A}_r.
\end{align}
By substituting \eqref{optimum_phases} in \eqref{received_power}, the maximum SNR as a function of $\mathcal{A}_r$ can be written as
\begin{align}
	\label{maximum_received_power}
	\gamma\left(\mathcal{A}_r\right) = \frac{P_{t}}{\sigma^2} 
	%\left(\frac{\lambda}{4\pi}\right)^4\frac{P_{t}G_{t}G_rG_{s}\left(\theta_{\rm{inc}}\right)G_{s}\left(\theta_{\rm{dep}}\right)}{d_t^2d_r^2}
	\left(\sum_{k \in \mathcal{A}_r}\left|h_{t_k}\right|\left|h_{r_k}\right|\right)^2.
\end{align}
We further note setting the impedance of the UCs dedicated for beamsteering in a way that perfect reflection is achieved with a phase response given by \eqref{optimum_phases} indicates that independent tuning of the UC amplitude and phase response can be achieved. Although this, in general, does not hold and a level of inter-dependency between the UC amplitude and phase response is expected \cite{Abeywickrama_TCOM_Practical_Phase_Shift}, there have been advanced literature designs considerably mitigating such coupling \cite{jia_broadband_2016}, \cite{Zhang2019}. Hence, we can view \eqref{maximum_received_power} as an upper bound of the maximum SNR that can be achieved with more advanced RIS designs

\subsection{Problem formulation}

\label{Problem_formulation}

There are two conflicting goals of the RIS operation: maximization of the SNR and maximization of the harvested power. To handle this multi-objective optimization, we consider the following two problem formulations: i) maximization of the SNR while guaranteeing a predetermined DC output power at the RIS and ii) maximization of the harvested power under a predetermined SNR requirement $\gamma_{0}$. 

\subsubsection*{Problem A: End-to-end SNR maximization}

This problem is formulated as
\begin{align}
	\label{Problem_1}
	\begin{array}{l l}
		\underset{\mathcal{A}_r}{\mathrm{maximize}} & \gamma\left(\mathcal{A}_{r}\right) \\
		\mathrm{subject\,\,to} & P_{\text{DC}}\left(\mathcal{A}_h\right)\geq P_{\text{RIS}}.
	\end{array}
\end{align}

Based on \eqref{DC_power}, \eqref{Harvested_power}, and \eqref{maximum_received_power}, \eqref{Problem_1} can be rewritten as

\begin{align}
	\label{Problem_1_rewritten}
	\begin{gathered}
		\underset{\mathcal{A}_r}{\mathrm{maximize}}\;\;\left(\sum_{k \in \mathcal{A}_r}\left|h_{t_k}\right|\left|h_{r_k}\right|\right)^2\\
	\mathrm{subject\,\,to} \;\; \frac{\frac{P_{\rm{max}}}{1+\exp\left(-a\left(\eta_{\rm{RF}}\
	P_{t}
	%left(\frac{\lambda}{4\pi}\right)^2	\frac{G_{t}G_{s}\left(\theta_{\rm{inc}}\right)}{d_t^2}
			\sum_{i \in \mathcal{A}_h}\left|h_{t_i}\right|^2-b\right)\right)}-\frac{P_{\rm{max}}}{1+e^{ab}}}{1-\frac{1}{1+e^{ab}}}\ge P_{\rm{RIS}}.
	\end{gathered}
\end{align}

\subsubsection*{Problem B: Harvested power maximization}

This problem is formulated as
\begin{align}
	\label{Problem_2}
	\begin{gathered}
		\underset{\mathcal{A}_h}{\mathrm{maximize}}\;\; P_{\rm{DC}}\left(\mathcal{A}_
		h\right)\\
		\mathrm{subject\,\,to} \;\; \gamma\left(\mathcal{A}_r\right) \ge \gamma_0,
	\end{gathered}
\end{align}

Based on \eqref{DC_power}, \eqref{Harvested_power}, and \eqref{maximum_received_power}, \eqref{Problem_2} can be rewritten as
\begin{align}
	\label{Problem_2_rewritten}
	\begin{gathered}
		\underset{\mathcal{A}_h}{\mathrm{maximize}}\;\;\sum_{i \in \mathcal{A}_h}\left|h_{t_i}\right|^2\\
		\text{subject to} \;\;
		\frac{P_{t}}{\sigma^2}
		%\left(\frac{\lambda}{4\pi}\right)^4\frac{P_{t}G_{t}G_rG_{s}\left(\theta_{\rm{inc}}\right)G_{s}\left(\theta_{\rm{dep}}\right)}{d_t^2d_r^2\sigma^2}
		\left(\sum_{k \in \mathcal{A}_r}\left|h_{t_k}\right|\left|h_{r_k}\right|\right)^2 \ge \gamma_0.
	\end{gathered}
\end{align}

The optimization problems \eqref{Problem_1_rewritten} and \eqref{Problem_2_rewritten} are combinatorial and a solution in closed form does not exist in the general case, where the channel gains $|h_{t_k}|$ and $|h_{r_k}|$ vary with $k$ due to multipath fading.
The combinatorial nature means that a brute-force search approach (exhaustive search) is required to find the optimal $\mathcal{A}_r$ and $\mathcal{A}_h$. By denoting the number of possible combinations to be examined as $N_{\rm{comb}}$, it holds that
\begin{align}
	N_{\rm{comb}}=\sum_{l=1}^{M_s-1}{M_s \choose l}=2^{M_s}-2.
\end{align} Thus, $N_{\rm{comb}}$ can be very large even for moderate $M_s$ and a brute-force approach is not viable in terms of execution time.\footnote{For instance, for $M_s=25$ it holds that $N_{\rm{comb}}=33554430$.} As a result, in Section~\ref{problem_solution} we provide sub-optimal policies for solving Problems A and B of substantially lower complexity.

\section{Proposed policies}

\label{problem_solution}

In this section, we first provide closed-form and low-complexity sub-optimal policies for solving Problems A and B. Subsequently, we prove that in the special case of free-space propagation conditions in the TX-RIS link, some of the proposed policies deliver the optimal solutions to the respective problems. 

Before we proceed, let us first denote the corresponding outputs $\mathcal{A}_r$ and $\mathcal{A}_h$ of the policies presented in this section for providing solutions to Problems A and B by $\mathcal{A}_r^{\left(s\right)}$ and $\mathcal{A}_h^{\left(s\right)}$, where $s \in \left\{A,B\right\}$. In addition, let us define function $\mathcal{L}\left(\cdot\right)$ that takes as argument channel gains and returns as output the UC indices that correspond to the particular gains.

\subsection{Problem A: End-to-end SNR maximization policy}

\label{Section_Problem_A_Solution}

 By directly inspecting \eqref{Problem_1_rewritten},  we observe that the utility function to be maximized depends on both the TX-RIS and RIS-RX channel gains. Hence, a natural choice of sub-optimal policies for solving Problem A involves channel-gain ordering of either the RIS-RX links, or product of the TX-RIS and RIS-RX links, or TX-RIS links, and selection of the highest number of the respective UCs for beamsteering for which the constraint of \eqref{Problem_1_rewritten} is satisfied. This reasoning gives rise to Algorithms A.1, A.2, and A.3 that are described as follows:
 
\begin{algorithm}[H]
	 \begin{small}
	\renewcommand{\thealgorithm}{A.1}
	\caption{Ordering of the RIS-RX link channel gains}
	\begin{algorithmic}[1]
		\renewcommand{\algorithmicrequire}{\textbf{Input:}}
		\renewcommand{\algorithmicensure}{\textbf{Output:}}
		%\REQUIRE in
		%\ENSURE  out
		%\\ \textit{Initialisation} :
		\STATE Arrange $\left|h_{r_m}\right|$, $m=1,2,...,M_s$, in descending order, i.e. $\left|h_{r_{\left(1\right)}}\right|\ge \left|h_{r_{\left(2\right)}}\right|\ge,...,\ge\left|h_{r_{\left(M_s\right)}}\right|$.Set iteration index $i=1$
		%\\ \textit{LOOP Process}
		\\\STATE\textbf{repeat} $\left\{\text{Loop}\right\}$\\
		\STATE\hphantom{~~~~}Set $	\mathcal{A}_r=\left\{\mathcal{L}\left(\left|h_{r_{\left(1\right)}}\right|,...,\left|h_{r_{\left(i\right)}}\right|\right)\right\}$, $\mathcal{A}_h=\left\{\mathcal{L}\left(\left|h_{r_{\left(i+1\right)}}\right|,...,\left|h_{r_{\left(M_s\right)}}\right|\right)\right\}$
		\\\STATE \textbf{until} Constraint in \eqref{Problem_1_rewritten} is not satisfied for $i=i_{\rm{stop}}^{\left(A\right)}$
		%\ENDFOR
		\\\STATE \textbf{Output} $\mathcal{A}_r^{\left(A\right)}=\left\{\mathcal{L}\left(\left|h_{r_{\left(1\right)}}\right|,...,\left|h_{r_{\left(i_{\rm{stop}}^{\left(A\right)}-1\right)}}\right|\right)\right\}$, $\mathcal{A}_h^{\left(A\right)}=\left\{\mathcal{L}\left(\left|h_{r_{\left(i_{\rm{stop}}^{\left(A\right)}\right)}}\right|,...,\left|h_{r_{\left(M_s\right)}}\right|\right)\right\}$  
	\end{algorithmic}
\end{small}
\end{algorithm}

\begin{algorithm}[H]
	\begin{small}
	\renewcommand{\thealgorithm}{A.2}
	\caption{Ordering of the product of the TX-RIS and RIS-RX link channel gains}
	\begin{algorithmic}[1]
		\renewcommand{\algorithmicrequire}{\textbf{Input:}}
		\renewcommand{\algorithmicensure}{\textbf{Output:}}
		%\REQUIRE in
		%\ENSURE  out
		%\\ \textit{Initialisation} :
		\STATE Arrange $g_m=\left|h_{t_m}\right|\left|h_{r_m}\right|$, $m=1,2,...,M_s$, in descending order, i.e. $g_{\left(1\right)}\ge g_{\left(2\right)}\ge,...,\ge g_{\left(M_s\right)}$. Set iteration index $i=1$
		%\\ \textit{LOOP Process}
		\\\STATE\textbf{repeat} $\left\{\text{Loop}\right\}$\\
		\STATE\hphantom{~~~~}Set $	\mathcal{A}_r=\left\{g_{\left(1\right)},..., g_{\left(i\right)} \right\}$, $\mathcal{A}_h=\left\{g_{\left(i+1\right)},..., g_{\left(M_s\right)}\right\}$
		\\\STATE \textbf{until} Constraint in \eqref{Problem_1_rewritten} is not satisfied for $i=i_{\rm{stop}}^{\left(A\right)}$
		%\ENDFOR
		\\\STATE \textbf{Output} $\mathcal{A}_r^{\left(A\right)}=\left\{\mathcal{L}\left(g_{\left(1\right)},...,g_{\left(i_{\rm{stop}}^{\left(A\right)}-1\right)}\right)\right\}$, $\mathcal{A}_h^{\left(A\right)}=\left\{\mathcal{L}\left(g_{i_{\left(\rm{stop}\right)}},...,g_{\left(M_s\right)}\right)\right\}$  
	\end{algorithmic}
\end{small} 
\end{algorithm}

\begin{algorithm}[H]
	\begin{small}
	\renewcommand{\thealgorithm}{A.3}
	\caption{Ordering of the TX-RIS link channel gains}
	\begin{algorithmic}[1]
		\renewcommand{\algorithmicrequire}{\textbf{Input:}}
		\renewcommand{\algorithmicensure}{\textbf{Output:}}
		%\REQUIRE in
		%\ENSURE  out
		%\\ \textit{Initialisation} :
		\STATE Arrange $\left|h_{t_m}\right|$, $m=1,2,...,M_s$, in descending order, i.e. $\left|h_{t_{\left(1\right)}}\right|\ge \left|h_{t_{\left(2\right)}}\right|\ge,...,\ge\left|h_{t_{\left(M_s\right)}}\right|$.Set iteration index $i=1$
		%\\ \textit{LOOP Process}
		\\\STATE \textbf{repeat} $\left\{\text{Loop}\right\}$\\
		\STATE\hphantom{~~~~}Set $	\mathcal{A}_r=\left\{\mathcal{L}\left(\left|h_{t_{\left(1\right)}}\right|,...,\left|h_{t_{\left(i\right)}}\right|\right)\right\}$, $\mathcal{A}_h=\left\{\mathcal{L}\left(\left|h_{t_{\left(i+1\right)}}\right|,...,\left|h_{t_{\left(M_s\right)}}\right|\right)\right\}$
		\\\STATE \textbf{until} Constraint in \eqref{Problem_1_rewritten} is not satisfied for $i=i_{\rm{stop}}^{\left(A\right)}$
		%\ENDFOR
		\\\STATE \textbf{Output} $\mathcal{A}_r^{\left(A\right)}=\left\{\mathcal{L}\left(\left|h_{t_{\left(1\right)}}\right|,...,\left|h_{t_{\left(i_{\rm{stop}}^{\left(A\right)}-1\right)}}\right|\right)\right\}$, $\mathcal{A}_h^{\left(A\right)}=\left\{\mathcal{L}\left(\left|h_{t_{\left(i_{\rm{stop}}^{\left(A\right)}\right)}}\right|,...,\left|h_{t_{\left(M_s\right)}}\right|\right)\right\}$  
	\end{algorithmic}
\end{small}
\end{algorithm}

In addition to Algorithms A.1, A.2, and A.3 that target the utility function of \eqref{Problem_1_rewritten}, it is also reasonable to target the constraint of \eqref{Problem_1_rewritten}. To this end, a rational approach is to maximize the number of UCs participating in beamsteering by minimizing the number of UCs dedicated for energy harvesting. The latter is achieved by dedicating for harvesting the UCs associated with the largest values of $\left|h_{t_m}\right|$, according to the proposed Algorithm A.4 that is described as follows:

\begin{algorithm}
	\begin{small}
	\renewcommand{\thealgorithm}{A.4}
	\caption{Ordering of the TX-RIS link channel gains}
	\begin{algorithmic}[1]
		\renewcommand{\algorithmicrequire}{\textbf{Input:}}
		\renewcommand{\algorithmicensure}{\textbf{Output:}}
		%\REQUIRE in
		%\ENSURE  out
		%\\ \textit{Initialisation} :
		\STATE Arrange $\left|h_{t_m}\right|$, $m=1,2,...,M_s$, in descending order, i.e. $\left|h_{t_{\left(1\right)}}\right|\ge \left|h_{t_{\left(2\right)}}\right|\ge,...,\ge\left|h_{t_{\left(M_s\right)}}\right|$. Set iteration index $i=1$
		%\\ \textit{LOOP Process}
		\\\STATE \textbf{repeat} $\left\{\text{Loop}\right\}$\\
		\STATE\hphantom{~~~~}Set $	\mathcal{A}_r=\left\{\mathcal{L}\left(\left|h_{t_{\left(i+1\right)}}\right|,...,\left|h_{t_{\left(M_s\right)}}\right|\right)\right\}$, $\mathcal{A}_h=\left\{\mathcal{L}\left(\left|h_{t_{\left(1\right)}}\right|,...,\left|h_{t_{\left(i\right)}}\right|\right)\right\}$
		\\\STATE \textbf{until} Constraint in \eqref{Problem_1_rewritten} is not satisfied for $i=i_{\rm{stop}}^{\left(A\right)}$
		%\ENDFOR
		\\\STATE \textbf{Output} $	\mathcal{A}_r^{\left(A\right)}=\left\{\mathcal{L}\left(\left|h_{t_{\left(i_{\rm{stop}}^{\left(A\right)}+1\right)}}\right|,...,\left|h_{t_{\left(M_s\right)}}\right|\right)\right\}$, $\mathcal{A}_h^{\left(A\right)}=\left\{\mathcal{L}\left(\left|h_{t_{\left(1\right)}}\right|,...,\left|h_{t_{\left(i_{\rm{stop}}^{\left(A\right)}\right)}}\right|\right)\right\}$
	\end{algorithmic}
\end{small}
\end{algorithm}

\subsubsection{Special case of equal-gain propagation in the TX-RIS link}

In this special case, we have $|h_{t_1}|=\ldots=|h_{t_{M_s}}|$, which means that the pathloss is the same between the TX and every UC. This condition is exactly satisfied in free-space far-field propagation, but can also be approximately occurring in other scenarios with a dominant line-of-sight (LoS) path (i.e. in Rician channels with large K-factor). Furthermore, we focus on the TX-RIS link for this case study since for fixed positions of the TX and RIS the TX-RIS link is expected to be quasi-static in nature and, consequently, close to the ideal case of equal-gain propagation. This would arise in likely scenarios where the TX is a base station and the RIS is mounted on a fixed structure, such as a building.
In this special case, we can prove the following proposition.

\begin{proposition}
	\label{Proposition_1}
	Under equal-gain propagation in the TX-RIS link with $|h_{t_1}|^2=\ldots=|h_{t_{M_s}}|^2 = \beta$, Algorithms A.1 and A.2 deliver the optimal selection of $\mathcal{A}_r^{\left(A\right)}$ and $\mathcal{A}_h^{\left(A\right)}$ with
	\begin{align}
		\label{i_stop_Problem_A_free_space}
		i_{\rm{stop}}^{\left(A\right)}=M_s-1-\left\lceil{\frac{-\left(\frac{1}{a}\right)\log\left(\frac{P_{\rm{max}}}{P_{\rm{RIS}}\left(1-\frac{1}{1+e^{ab}}\right)+\frac{P_{\rm{max}}}{1+e^{ab}}}-1\right)+b}{\eta_{\rm{RF}}
		P_t \beta 
				% \left(\frac{\lambda}{4\pi}\right)^2\frac{P_{t}G_{t}G_{s}\left(\theta_{\rm{inc}}\right)}{d_t^2}
				}}\right\rceil,
	\end{align}
	where $\lceil\cdot\rceil$ is the ceiling function.
\end{proposition}
\begin{IEEEproof}
The proof is given in Appendix~\ref{app:prop1}.
\end{IEEEproof}

We note that the higher the deviation of the propagation conditions in the TX-RIS link is with respect to equal-gain propagation, the higher the performance gap between the brute-force approach and Algorithms A.1 and A.2 is expected to be. This will be verified in the numerical results of Section~\ref{Numerical_results}.

\subsection{Problem B: Harvested power maximization policy}

\label{Section_Problem_B_Solution}

A direct inspection of \eqref{Problem_2_rewritten} reveals that the utility function to be maximized depends only on the TX-RIS channel gains. Hence, a natural sub-optimal solution to Problem B involves channel-gain ordering of the TX-RIS links and selection of the highest number of the respective UCs for harvesting, satisfying the constraint of Problem B. This reasoning results to Algorithm B.1 that is described as follows:

\begin{algorithm}[H]
	\begin{small}
	\renewcommand{\thealgorithm}{B.1}
	\caption{Ordering of the TX-RIS link channel gains}
	\begin{algorithmic}[1]
		\renewcommand{\algorithmicrequire}{\textbf{Input:}}
		\renewcommand{\algorithmicensure}{\textbf{Output:}}
		%\REQUIRE in
		%\ENSURE  out
		%\\ \textit{Initialisation} :
		\STATE Arrange $\left|h_{t_m}\right|$, $m=1,2,...,M_s$, in descending order, i.e. $\left|h_{t_{\left(1\right)}}\right|\ge \left|h_{t_{\left(2\right)}}\right|\ge,...,\ge\left|h_{t_{\left(M_s\right)}}\right|$. Set iteration index $i=1$
		%\\ \textit{LOOP Process}
		\\\STATE \textbf{repeat} $\left\{\text{Loop}\right\}$\\
		\STATE\hphantom{~~~~}Set $	\mathcal{A}_r=\left\{\mathcal{L}\left(\left|h_{t_{\left(i+1\right)}}\right|,...,\left|h_{t_{\left(M_s\right)}}\right|\right)\right\}$, $\mathcal{A}_h=\left\{\mathcal{L}\left(\left|h_{t_{\left(1\right)}}\right|,...,\left|h_{t_{\left(i\right)}}\right|\right)\right\}$
		\\\STATE \textbf{until} Constraint in \eqref{Problem_2_rewritten} is not satisfied for $i=i_{\rm{stop}}^{\left(B\right)}$
		%\ENDFOR
		\\\STATE \textbf{Output} $	\mathcal{A}_r^{\left(B\right)}=\left\{\mathcal{L}\left(\left|h_{t_{\left(i_{\rm{stop}}^{\left(B\right)}\right)}}\right|,...,\left|h_{t_{\left(M_s\right)}}\right|\right)\right\}$, $\mathcal{A}_h^{\left(B\right)}=\left\{\mathcal{L}\left(\left|h_{t_{\left(1\right)}}\right|,...,\left|h_{t_{\left(i_{\rm{stop}}^{\left(B\right)}-1\right)}}\right|\right)\right\}$
	\end{algorithmic}
\end{small}
\end{algorithm}

In addition to Algorithm B.1 that targets the utility function of \eqref{Problem_2_rewritten}, it is also reasonable to target the constraint of \eqref{Problem_2_rewritten}. On this account, a rational approach is to increase the number of UCs participating in energy harvesting by reducing the number of UCs dedicated for reflection. The latter is achieved by dedicating for reflection the UCs associated with the largest values of either $\left|h_{r_m}\right|$, or $\left|h_{t_m}\right|\left|h_{r_m}\right|$, or $\left|h_{t_m}\right|$, according to the proposed Algorithms B.2, B.3, and B.4, respectively,  that are described as follows:

\begin{algorithm}[H]
	\begin{small}
	\renewcommand{\thealgorithm}{B.2}
	\caption{Ordering of the RIS-RX link channel gains}
	\begin{algorithmic}[1]
		\renewcommand{\algorithmicrequire}{\textbf{Input:}}
		\renewcommand{\algorithmicensure}{\textbf{Output:}}
		%\REQUIRE in
		%\ENSURE  out
		%\\ \textit{Initialisation} :
		\STATE Arrange $\left|h_{r_m}\right|$, $m=1,2,...,M_s$, in descending order, i.e. $\left|h_{r_{\left(1\right)}}\right|\ge \left|h_{r_{\left(2\right)}}\right|\ge,...,\ge\left|h_{r_{\left(M_s\right)}}\right|$.Set iteration index $i=1$
		%\\ \textit{LOOP Process}
		\\\STATE\textbf{repeat} $\left\{\text{Loop}\right\}$\\
		\STATE\hphantom{~~~~}Set $	\mathcal{A}_r=\left\{\mathcal{L}\left(\left|h_{r_{\left(1\right)}}\right|,...,\left|h_{r_{\left(i\right)}}\right|\right)\right\}$, $\mathcal{A}_h=\left\{\mathcal{L}\left(\left|h_{r_{\left(i+1\right)}}\right|,...,\left|h_{r_{\left(M_s\right)}}\right|\right)\right\}$
		\\\STATE \textbf{until} Constraint in \eqref{Problem_2_rewritten} is satisfied for $i=i_{\rm{stop}}^{\left(B\right)}$
		%\ENDFOR
		\\\STATE \textbf{Output} $\mathcal{A}_r^{\left(B\right)}=\left\{\mathcal{L}\left(\left|h_{r_{\left(1\right)}}\right|,...,\left|h_{r_{\left(i_{\rm{stop}}^{\left(B\right)}\right)}}\right|\right)\right\}$, $\mathcal{A}_h^{\left(B\right)}=\left\{\mathcal{L}\left(\left|h_{r_{\left(i_{\rm{stop}}^{\left(B\right)}+1\right)}}\right|,...,\left|h_{r_{\left(M_s\right)}}\right|\right)\right\}$  
	\end{algorithmic}
\end{small} 
\end{algorithm}

\begin{algorithm}[H]
	\begin{small}
	\renewcommand{\thealgorithm}{B.3}
	\caption{Ordering of the product of the TX-RIS and RIS-RX link channel gains}
	\begin{algorithmic}[1]
		\renewcommand{\algorithmicrequire}{\textbf{Input:}}
		\renewcommand{\algorithmicensure}{\textbf{Output:}}
		%\REQUIRE in
		%\ENSURE  out
		%\\ \textit{Initialisation} :
		\STATE Arrange $g_m=\left|h_{t_m}\right|\left|h_{r_m}\right|$, $m=1,2,...,M_s$, in descending order, i.e. $g_{\left(1\right)}\ge g_{\left(2\right)}\ge,...,\ge g_{\left(M_s\right)}$. Set iteration index $i=1$
		%\\ \textit{LOOP Process}
		\\\STATE\textbf{repeat} $\left\{\text{Loop}\right\}$\\
		\STATE\hphantom{~~~~}Set $	\mathcal{A}_r=\left\{g_{\left(1\right)},..., g_{\left(i\right)} \right\}$, $\mathcal{A}_h=\left\{g_{\left(i+1\right)},..., g_{\left(M_s\right)}\right\}$
		\\\STATE \textbf{until} Constraint in \eqref{Problem_2_rewritten} is satisfied for $i=i_{\rm{stop}}^{\left(B\right)}$
		%\ENDFOR
		\\\STATE \textbf{Output} $\mathcal{A}_r^{\left(B\right)}=\left\{\mathcal{L}\left(g_{\left(1\right)},...,g_{\left(i_{\rm{stop}}^{\left(B\right)}\right)}\right)\right\}$, $\mathcal{A}_h^{\left(B\right)}=\left\{\mathcal{L}\left(g_{i_{\left(\rm{stop}+1\right)}},...,g_{\left(M_s\right)}\right)\right\}$  
	\end{algorithmic}
\end{small} 
\end{algorithm}

\begin{algorithm}[H]
	\begin{small}
	\renewcommand{\thealgorithm}{B.4}
	\caption{Ordering of the TX-RIS link channel gains}
	\begin{algorithmic}[1]
		\renewcommand{\algorithmicrequire}{\textbf{Input:}}
		\renewcommand{\algorithmicensure}{\textbf{Output:}}
		%\REQUIRE in
		%\ENSURE  out
		%\\ \textit{Initialisation} :
		\STATE Arrange $\left|h_{t_m}\right|$, $m=1,2,...,M_s$, in descending order, i.e. $\left|h_{t_{\left(1\right)}}\right|\ge \left|h_{t_{\left(2\right)}}\right|\ge,...,\ge\left|h_{t_{\left(M_s\right)}}\right|$.Set iteration index $i=1$
		%\\ \textit{LOOP Process}
		\\\STATE \textbf{repeat} $\left\{\text{Loop}\right\}$\\
		\STATE\hphantom{~~~~}Set $	\mathcal{A}_r=\left\{\mathcal{L}\left(\left|h_{t_{\left(1\right)}}\right|,...,\left|h_{t_{\left(i\right)}}\right|\right)\right\}$, $\mathcal{A}_h=\left\{\mathcal{L}\left(\left|h_{t_{\left(i+1\right)}}\right|,...,\left|h_{t_{\left(M_s\right)}}\right|\right)\right\}$
		\\\STATE \textbf{until} Constraint in \eqref{Problem_2_rewritten} is satisfied for $i=i_{\rm{stop}}^{\left(B\right)}$
		%\ENDFOR
		\\\STATE \textbf{Output} $\mathcal{A}_r^{\left(B\right)}=\left\{\mathcal{L}\left(\left|h_{t_{\left(1\right)}}\right|,...,\left|h_{t_{\left(i_{\rm{stop}}^{\left(B\right)}\right)}}\right|\right)\right\}$, $\mathcal{A}_h^{\left(B\right)}=\left\{\mathcal{L}\left(\left|h_{t_{\left(i_{\rm{stop}}^{\left(B\right)}+1\right)}}\right|,...,\left|h_{t_{\left(M_s\right)}}\right|\right)\right\}$  
	\end{algorithmic}
\end{small}
\end{algorithm}

\subsubsection{Special case of equal-gain propagation in the TX-RIS link}

For this special case, still characterized by $|h_{t_1}|=\ldots=|h_{t_{M_s}}|$, we can prove the following result.

\begin{proposition}
	Under equal-gain propagation in the TX-RIS link with $|h_{t_1}|^2=\ldots=|h_{t_{M_s}}|^2 = \beta$, Algorithms B.2 and B.3 deliver the optimal allocation for $\mathcal{A}_r^{\left(B\right)}$ and $\mathcal{A}_h^{\left(B\right)}$.
	\end{proposition}
\begin{IEEEproof}
The proof is obtained straightforwardly by considering that $P_{\rm{DC}}$ is maximized when the number of UCs acting as beamformers that satisfy the constraint of \eqref{Problem_2_rewritten}, is minimized. For $\left|h_{t_m}\right|=\sqrt{\beta}$, $m=1,2,...,M_s$, this is achieved by using for beamsteering the UCs associated with the largest values of $\left|h_{r_m}\right|=\sqrt{\beta}$, $m=1,2,...,M_s$, as described by Algorithm B.2. Furthermore, this optimal solution can also be reached by Algorithm B.3 due to fact that for $\left|h_{t_m}\right|=\sqrt{\beta}$, Algorithm B.3 degenerates to Algorithm B.2.
\end{IEEEproof}

We note that again, as it was pointed out in Problem A, the higher the deviation of the propagation conditions in the TX-RIS link is with respect to equal-gain propagation, the higher the achieved performance gap between the brute force approach and Algorithms B.2 and B.3 is anticipated. This trend is verified in the numerical results presented in the next section.

\section{Numerical Results and Discussion}

\label{Numerical_results}

In this section, we first provide a case study for the propagation conditions in the TX-RIS and RIS-RX links by incorporating the well-known Rician fading model. Subsequently, based on this channel model, we present numerical results related to the two considered optimization problems and the proposed solutions, which verify our claims. Finally, we conclude this section by providing a discussion, based on a plausible study, regarding the possibility for the involved RIS electronics to exhibit the necessary low power consumption values that are needed for autonomous RIS operation.

\subsection{Case study: Rician fading}

We assume that both the TX-RIS and RIS-RX links are subject to uncorrelated\footnote{This may approximately hold for only for $d_x$ and $d_y$ equal to half wavelength \cite{Emil_Correlation_RISs}. However, this does not limit the generality of our framework since the  correlation among the channels corresponding to the individual UCs can be incorporated into the framework according to an existing model that describes the relation between channel correlation and $d_x$, $d_y$ \cite{Emil_Correlation_RISs}.} Rician fading, with corresponding K-factors denoted by $K_1$ and $K_2$, respectively. Such a distribution is justified by the elevated position of an RIS in practical scenarios with distinct LoS paths and diffuse multipaths for both the TX-RIS and RIS-RX channels. In addition, the suitability of the Rician distribution is supported by channel measurements in both sub-6 GHz and mmWave bands \cite{Greenstein_2009}, \cite{Hanzo_MmWave_Channel_Model}. Furthermore, as far as the radiation pattern of each UC is concerned, we consider that each UC is an electrically-small low-gain element with a cosine gain pattern, with respect to the azimuth angle $\theta$, expressed as
\begin{align}
	\label{RIS_element_gain}
	G_{s}\left(\theta\right)= 4{\cos\left(\theta\right)}, \quad 0\le\theta<\pi/2.
\end{align}This model is supported by the measurements \cite{Tang_measurements_2021}. Finally, by $G_t$ and $G_r$, we denote the gains of the TX and RX antennas in the directions of the RIS. Based on these and considering for simplification a free-space propagation based path-loss exponent model\footnote{In cases of dominant LoS components, the actual path-loss exponent is expected to be close to the free-space propagation one.}, it holds that
\begin{align}
	\label{Rician_fading}
	\begin{gathered}
		{\bf{h}}_{\rm{t}}=\sqrt{\left(\frac{\lambda}{4\pi}\right)^2
			\frac{G_{t}G_{s}\left(\theta_{\rm{inc}}\right)}{d_t^2}} \left[e^{j\frac{2\pi}{\lambda}d_{t_1}}+m_1 \quad e^{j\frac{2\pi}{\lambda}d_{t_2}}+m_2 \quad 
		\cdots\quad e^{j\frac{2\pi}{\lambda}d_{t_{M_s}}}+m_{M_s} \right]\\
		{\bf{h}}_{\rm{r}}=\sqrt{\left(\frac{\lambda}{4\pi}\right)^2
			\frac{G_{r}G_{s}\left(\theta_{\rm{dep}}\right)}{d_r^2}} \left[e^{j\frac{2\pi}{\lambda}d_{r_1}}+p_1 \quad e^{j\frac{2\pi}{\lambda}d_{r_2}}+p_2 \quad 
		\cdots\quad e^{j\frac{2\pi}{\lambda}d_{r_{M_s}}}+p_{M_s} \right],
	\end{gathered}
\end{align}
where $\lambda$ is the wavelength and $d_{t_k}$, $d_{r_k}$, $k=1,2,....,M_s$, are the distances between the TX and the center of the $k_{\rm{th}}$ UC and between the center of the $k_{\rm{th}}$ UC and the RX, respectively. Furthermore, $\theta_{\rm{inc}}$ and $\theta_{\rm{dep}}$ denote the incident angle on the RIS and departure angle from the RIS of the dominant LoS component, respectively.  In addition, $m_k\in\mathcal{CN}\left(0,\sigma_t^2\right)$ and $p_k\in\mathcal{CN}\left(0,\sigma_r^2\right)$ represent the multipath complex envelopes of the Rayleigh fading describing the diffuse scattering in the TX-RIS and RIS-RX links, respectively. Hence, for the mentioned Rician $K_1$ and $K_2$ factors, it holds
\begin{align}
	K_1=\frac{1}{\sigma_t^2}, \quad K_2=\frac{1}{\sigma_r^2}.
\end{align}

\subsection{Results}

\label{Numerical_results_subsection}

For the numerical results, we consider the parameter values presented in Table~\ref{Parameter_values}.\footnote{In contrast to a linear structure, the considered $\lambda/2$ distance of adjacent UCs located in the same axis does not eliminate spatial correlation due to the RIS planar structure and the resulting $\frac{\lambda}{2}\sqrt{2}$ distance of adjacent UCs belonging to the different axes. However, due to the fact that such a correlation is small for $d_x=d_y=\lambda/2$ \cite{Emil_Correlation_RISs} and mainly affect the channel estimation and not the link budget, we ignore it in this work and consider uncorrelated Rician fading.} According to the particular values, it holds that $P_{d}^{\rm{avg}}=\alpha p_r P_{\rm{dynamic}}=8$ $\upmu$W.

\begin{table}[h]
	\caption{Parameter values used in the simulation.\label{Parameter_values}} % title of Table
	\centering % used for centering table
	\scalebox{0.8}{
		\begin{tabular}{| c | c | c | c | } % centered columns (4 columns)
			\hline
			Parameter & Value & Parameter & Value\\[0.5ex]
			\hline
			\hline
			$f$& $28$ GHz& $d_x$, $d_y$& $\lambda/2$ \\[0.5ex] 
			\hline
			$P_{t}$& $1$ W & $\sigma_r^2$& $0.3$ \\ [0.5ex]
			\hline
			$G_t$& 40 dBi & $G_r$& 22 dBi \\ [0.5ex]
			\hline
			$d_t$& 17 m& $d_r$ &  20 m\\ [0.5ex]
			\hline
			$\theta_{\rm{inc}}$& $45^{\circ}$ & $\theta_{\rm{dep}}$ & $60^{\circ}$ \\ [0.5ex]
			\hline
			$\mathcal{F}_{{\rm{dB}}}$& $10$ dB & $a$ & $120$\\[0.5ex]
			\hline
			$b$& $10^{-3}$ & $P_{\rm{max}}$ & $20$ mW \\[0.5ex]
			\hline
			$W$& 1 GHz & $\eta_{\rm{RF}}$& 0.5 \\[0.5ex]
			\hline
			$P_{\rm{static}}$ & 2 $\upmu$W & $\alpha$ & 0.8 \\[0.5ex]
			\hline
			$p_r$ & $10^{-3}$ & $P_{\rm{dynamic}}$ & 10 mW \\[0.5ex]
			\hline
			%$?_r=2$ &$?_r=3$& $?_r=2$ &$?_r=3$\\ [0.5ex]
			%\multicolumn{2}{|c|}{Proposed selection}& \multicolumn{2}{|c|}{Capacity-based selection}\\ [0.5ex]
			%Proposed selection&Capacity-based selection \\ [0.5ex]
	\end{tabular}}
\end{table}

\begin{figure}
	%\label{optimal_SNR_distance_vs_power_consumption}
	\centering
	{\includegraphics[width=6in, height=3in]{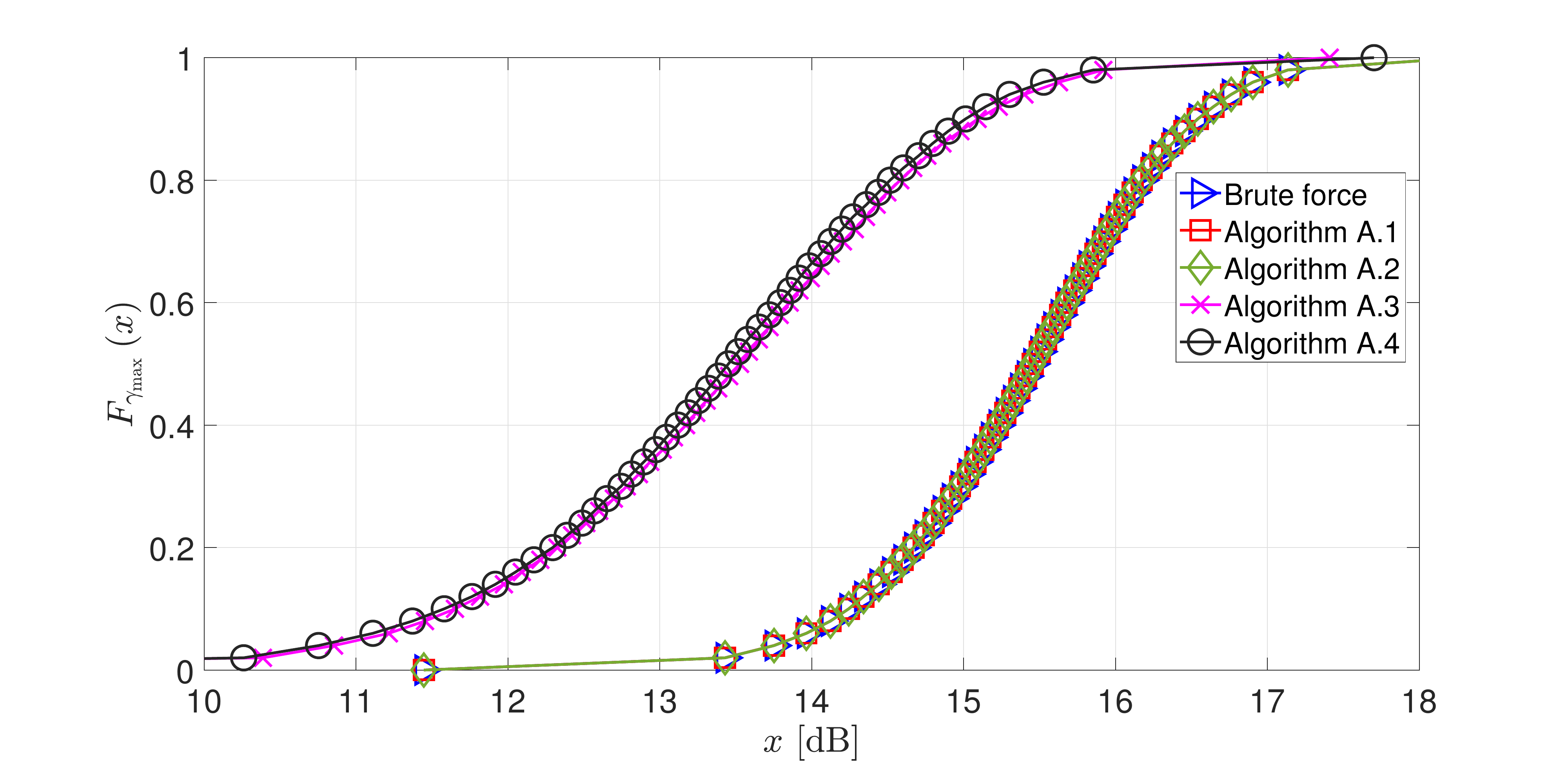}}\\
	(a) $\sigma_t^2=0$ ($K_1 \to \infty$).
	{\includegraphics[width=6in, height=3in]{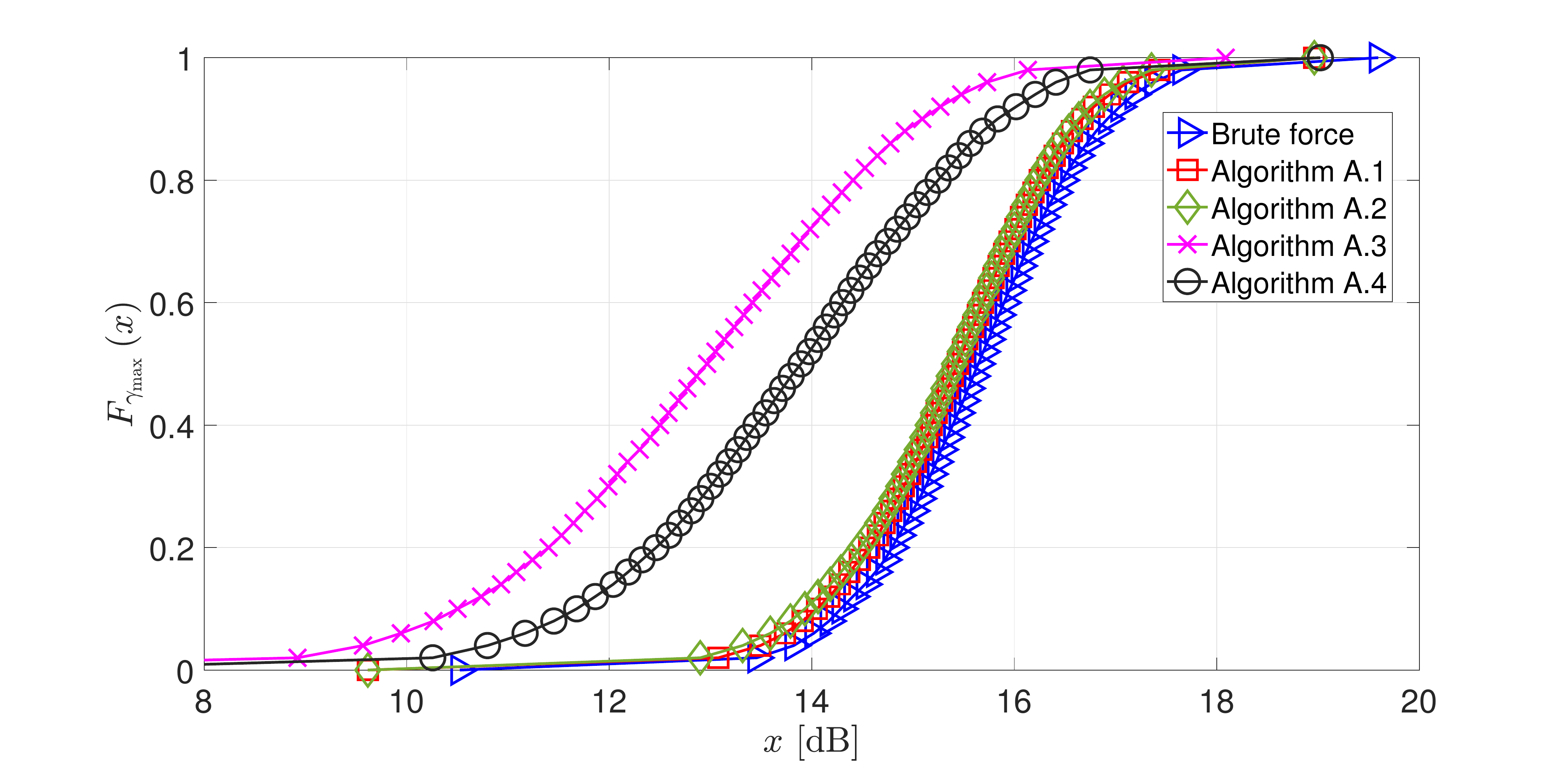}}\\
	(b) $\sigma_t^2=0.01$ ($K_1=100$).
	{\includegraphics[width=6in, height=3in]{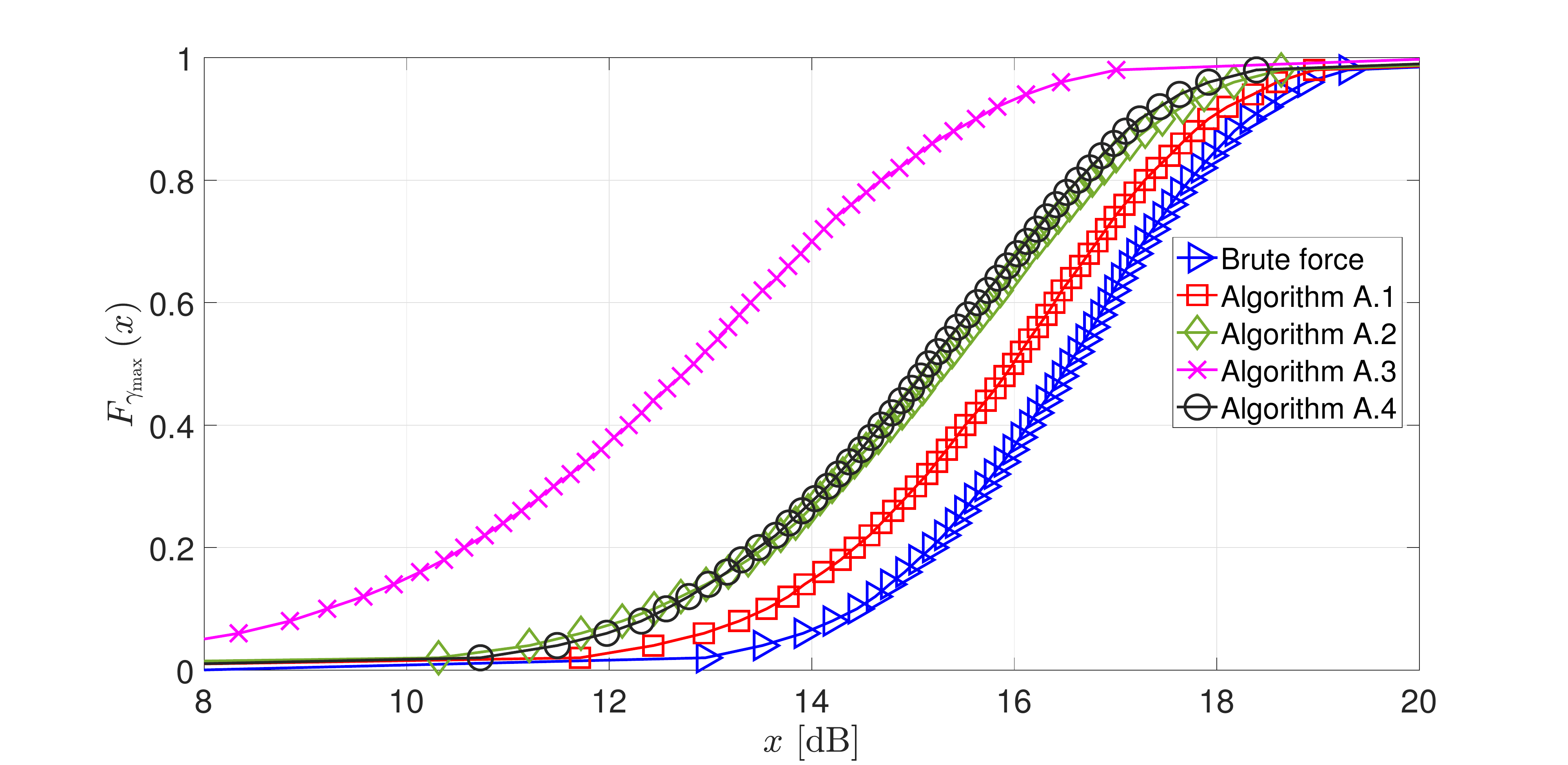}}\\
	(c) $\sigma_t^2=0.1$ ($K_1=10$).
	\caption{$F_{\gamma_{\rm{max}}}\left(x\right)$ vs. $x$ for $M_x=5$ and $M_y=2$ ($M_s=10$).}
	%\hrulefill
	\label{Problem_A_CDF}	
\end{figure}

\subsubsection{Problem A: End-to-end SNR maximization policy}

Regarding Problem A, our first goal is to examine how close the performance of the proposed Algorithms A.1, A.2, A.3, and A.4 is to the corresponding brute-force approach for channel conditions in the TX-RIS link that range from free-space propagation to notable scattering. Towards this, in Fig.~\ref{Problem_A_CDF} we depict the cumulative density function of $\gamma\left(\mathcal{A}_r^{\left(A\right)}\right)$, denoted by $F_{\gamma_{\rm{max}}}\left(x\right)$ and obtained by Monte-Carlo simulations for the brute-force approach and the proposed algorithms. As we observe, Algorithms A.1 and A.2 result in the same performance as the one of the brute-force approach for $\sigma_t^2=0$ and the performance gap increases with increasing $\sigma_t^2$, which validates our remark at the end of Section~\ref{Section_Problem_A_Solution}. However, we observe that even in the $\sigma_t^2=0.1$ case, which is a relatively high value, especially if the TX-RIS channel is quasi-static, the performance of Algorithm A.1 remains relatively close to the corresponding one of brute force.

\begin{figure}
	%\label{optimal_SNR_distance_vs_power_consumption}
	\centering
	{\includegraphics[width=6in, height=3in]{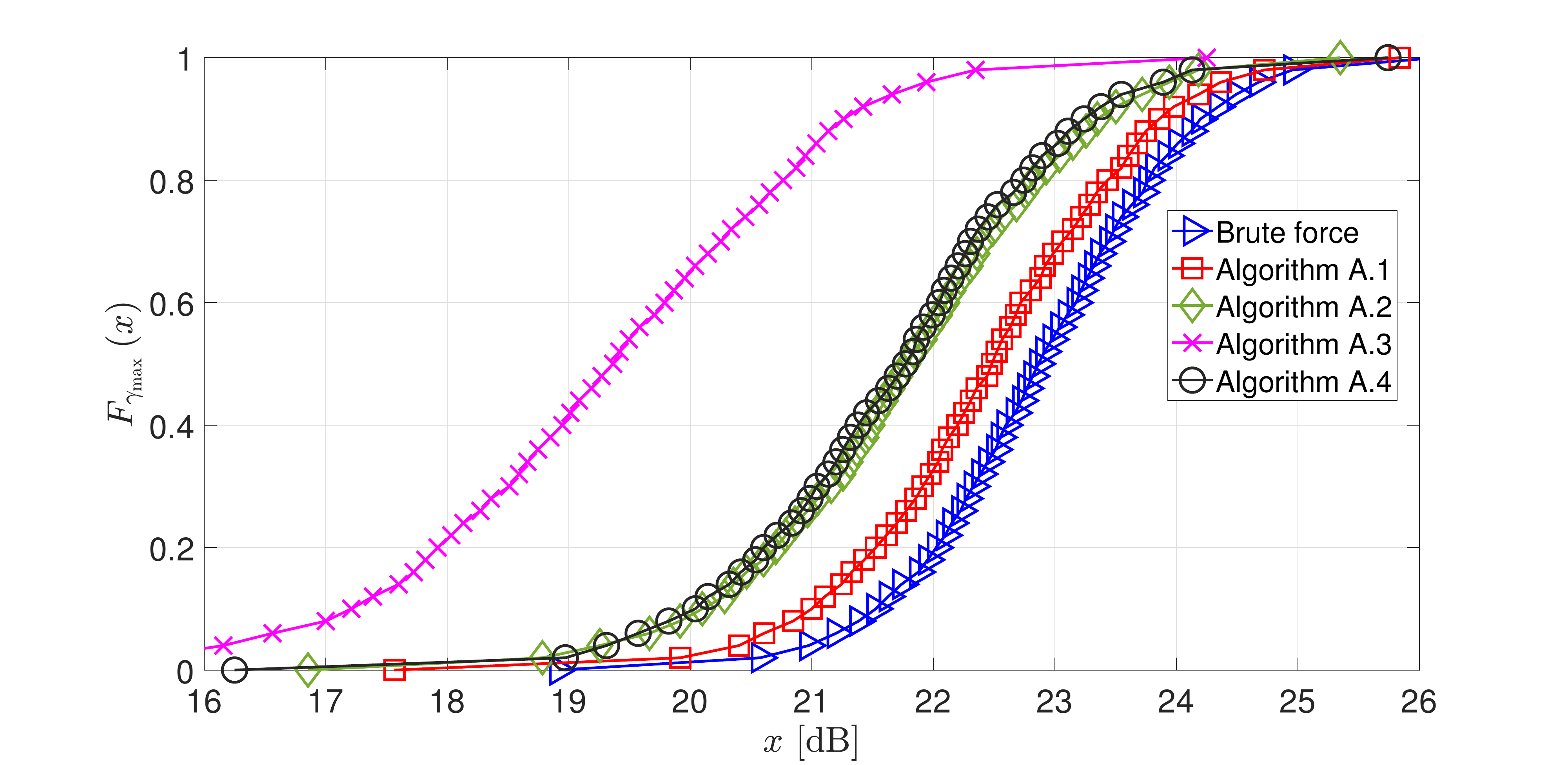}}
	\caption{$F_{\gamma_{\rm{max}}}\left(x\right)$ vs. $x$ for $M_x=5$, $M_y=4$ ($M_s=20$), and $\sigma_t^2=0.1$ ($K_1=10$).}
	%\hrulefill
	\label{Problem_1_variance_0_1}	
\end{figure}

Our second goal is to examine whether the relative performance gap of the proposed algorithms with respect to brute force increases for increasing $M_s$. Toward this end, in Fig.~\ref{Problem_1_variance_0_1} we depict $F_{\gamma_{\rm{max}}}\left(x\right)$ for $M_x=5$, $M_y=4$ ($M_s=20$), and $\sigma_t^2=0.1$ ($K_1=10$). Comparing the plots of Fig.~\ref{Problem_1_variance_0_1} with the ones of Fig.~\ref{Problem_A_CDF}-(c), where the latter refers to the $M_x=5$, $M_y=2$ ($M_s=10$) case, we observe that the performance gap of the algorithms with respect to brute force is almost constant. This indicates that it does not depend on $M_s$. This is an important finding, showing that the proposed simple algorithms that are based on channel-gain ordering can result in performance that is not far from the optimal one obtained by brute-force optimal, regardless of $M_s$. Such outcomes are further substantiated by Table~\ref{Problem_1_average_SNR_sigma_t_0_1}, which presents the average value of $\gamma(\mathcal{A}_r^{\left(A\right)})$, denoted by $\bar\gamma(\mathcal{A}_r^{\left(A\right)})$, for 4 values of $M_x$, $M_y$. As Table~\ref{Problem_1_average_SNR_sigma_t_0_1} reveals, the performance gap among the considered algorithms remains almost constant as $M_s$ increases. For Algorithm A.1, which results in the best performance among the sub-optimal algorithms, its performance gap with respect to brute force is within 0.4-0.5 dB.

\begin{table}[h]
	\begin{small}
		\begin{center}
			\caption{$\bar\gamma\left(\mathcal{A}_r^{\left(A\right)}\right)$ [dB] for $\sigma_t^2=0.1$ ($K_1=10$).	\label{Problem_1_average_SNR_sigma_t_0_1}}
			
			\begin{tabular}{|c|c|c|c|c|}
				\hline
				\multirow{2}{*}{Algorithm} & \multicolumn{4}{c|}{$M_x$, $M_y$} \\ \cline{2-5} 
				& 5, 2 ($M_s=10$)    & 4, 3 ($M_s=12$)  & 5, 3 ($M_s=15$) & 5, 4 ($M_s=20$)   \\ \hline
				Brute force                & 16.7  & 18.4 & 20.4 & 23.0   \\ \hline
				A.1                        & 16.2  & 17.9 & 19.9 & 22.6 \\ \hline
				A.2                        & 15.5  & 17.2 & 19.3 & 21.9 \\ \hline
				A.3                        & 13.2  & 15.0   & 17.0   & 19.6 \\ \hline
				A.4                        & 15.4  & 17.1 & 19.2 & 21.9 \\ \hline
			\end{tabular}
		\end{center}
	\end{small}
\end{table}

Finally, we will explain why Algorithm A.1 provides the best performance among the proposed algorithms. This trend is counter-intuitive since it would be expected that the best performance is achieved by an algorithm that takes also into account the gain of the TX-RIS links since they are involved in both the end-to-end SNR and DC harvested power, according to \eqref{Problem_1_rewritten}. However, we can justify the resulting trend thinking that such an inter-dependency between the utility function and constraint in \eqref{Problem_1_rewritten} due to the TX-RIS channel gains would dictate the need for an algorithm that is not based on the particular gains. This way, such an inter-dependency is balanced. Otherwise, an algorithm with which channel-gain ordering of the TX-RIS link targets either the utility function or the constraint in \eqref{Problem_1_rewritten} can have a negative effect on the constraint or the utility function, respectively, due to the conflicting inter-dependency. The aforementioned rational is further substantiated by Table~\ref{Number_of_UCs_harvesting}, which presents the probability mass function of the random variables that represents the number of harvesting UCs for the brute-force approach and the proposed algorithms in the case where $\sigma_t^2=0.1$, $M_x=5$, and $M_y=4$. 

\begin{table}[h]
	\begin{small}
		\begin{center}
			\caption{Probability mass function of $M_h$ for $\sigma_t^2=0.1$, $M_x=5$, and $M_y=4$. \label{Number_of_UCs_harvesting}}
			
			\begin{tabular}{|c|c|c|c|c|c|c|c|c|c|c|c|c|}
				\hline
				\multirow{2}{*}{Algorithm} & \multicolumn{12}{c|}{$M_h$}                                                                     \\ \cline{2-13} 
				& 4     & 5      & 6     & 7     & 8     & 9     & 10    & 11    & 12    & 13    & 14    & 15    \\ \hline
				Brute force                & 0.001 & 0.005 & 0.218 & 0.425 & 0.246 & 0.054 & 0.009 & 0.001 & 0     & 0     & 0     & 0     \\ \hline
				A.1                        & 0     & 0.025  & 0.115 & 0.284 & 0.312 & 0.190  & 0.052 & 0.017 & 0.004 & 0     & 0     & 0     \\ \hline
				A.2                        & 0     & 0      & 0.007 & 0.052 & 0.186 & 0.307 & 0.271 & 0.121 & 0.046 & 0.010  & 0     & 0     \\ \hline
				A.3                        & 0     & 0      & 0     & 0     & 0     & 0.005 & 0.064 & 0.270  & 0.398 & 0.199 & 0.055 & 0.009 \\ \hline
				A.4                        & 0.013 & 0.236  & 0.518 & 0.209 & 0.020  & 0.003 & 0.001 & 0     & 0     & 0     & 0     & 0     \\ \hline
			\end{tabular}
		\end{center}
		
	\end{small}
\end{table}

As we observe from Table~\ref{Number_of_UCs_harvesting}, the number of harvesting UCs with the highest probability is smaller for Algorithm A.1 than the corresponding one of Algorithms A.2 and A.3. This occurs due to the fact that the latter two algorithms involve channel-gain ordering which incorporates the TX-RIS link. Hence, the UCs selected for beamsteering in the particular algorithms exhibit on average higher values of the TX-RIS channel gains than the ones in Algorithm A.1. This explains the fact that a smaller number of UCs on average is required for energy harvesting in Algorithm A.1, which allows higher flexibility for the maximization of the SNR. In contrast, Algorithm A.4 results in a smaller number of selected harvesting UCs than the one in Algorithm A.1 since the former incorporates channel-gain ordering of the TX-RIS link that targets the minimization of the number of UCs satisfying the constraint in \eqref{Problem_1_rewritten}. On the one hand, this minimizes the number of harvesting UCs but, on the other hand, it results in smaller, on average, channel gains for the UCs participating in beamsteering compared to Algorithm A.1. This explains why the balanced approach of Algorithm A.1 that does not involve UC selection based on the TX-RIS channel gains results in the best performance among the proposed algorithms. 

\subsubsection{Problem B: Harvested power maximization policy}

\begin{figure}
	%\label{optimal_SNR_distance_vs_power_consumption}
	\centering
	{\includegraphics[width=6in, height=3in]{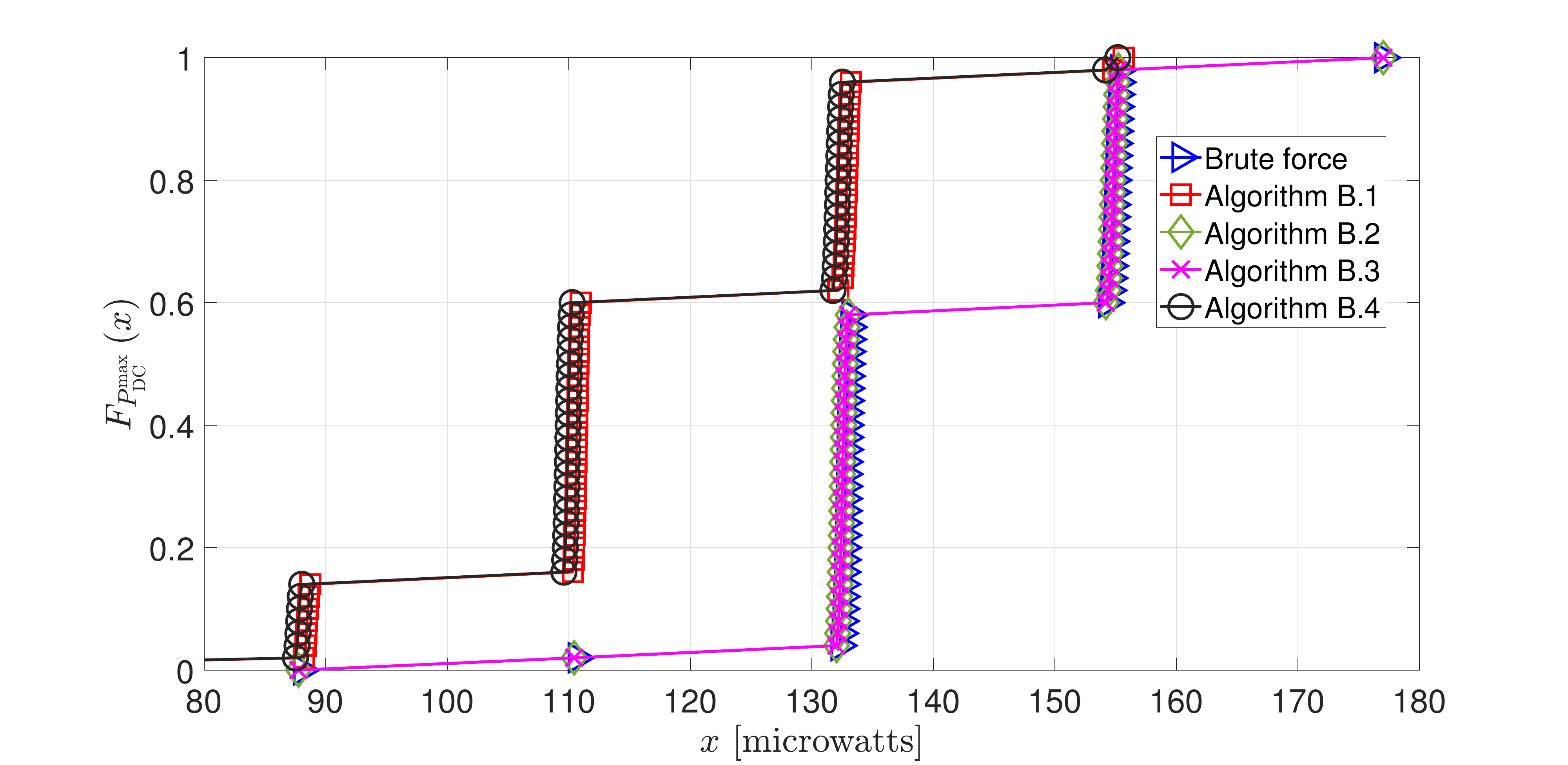}}\\
	(a) $\sigma_t^2=0$ ($K_1 \to \infty$).
	{\includegraphics[width=6in, height=3in]{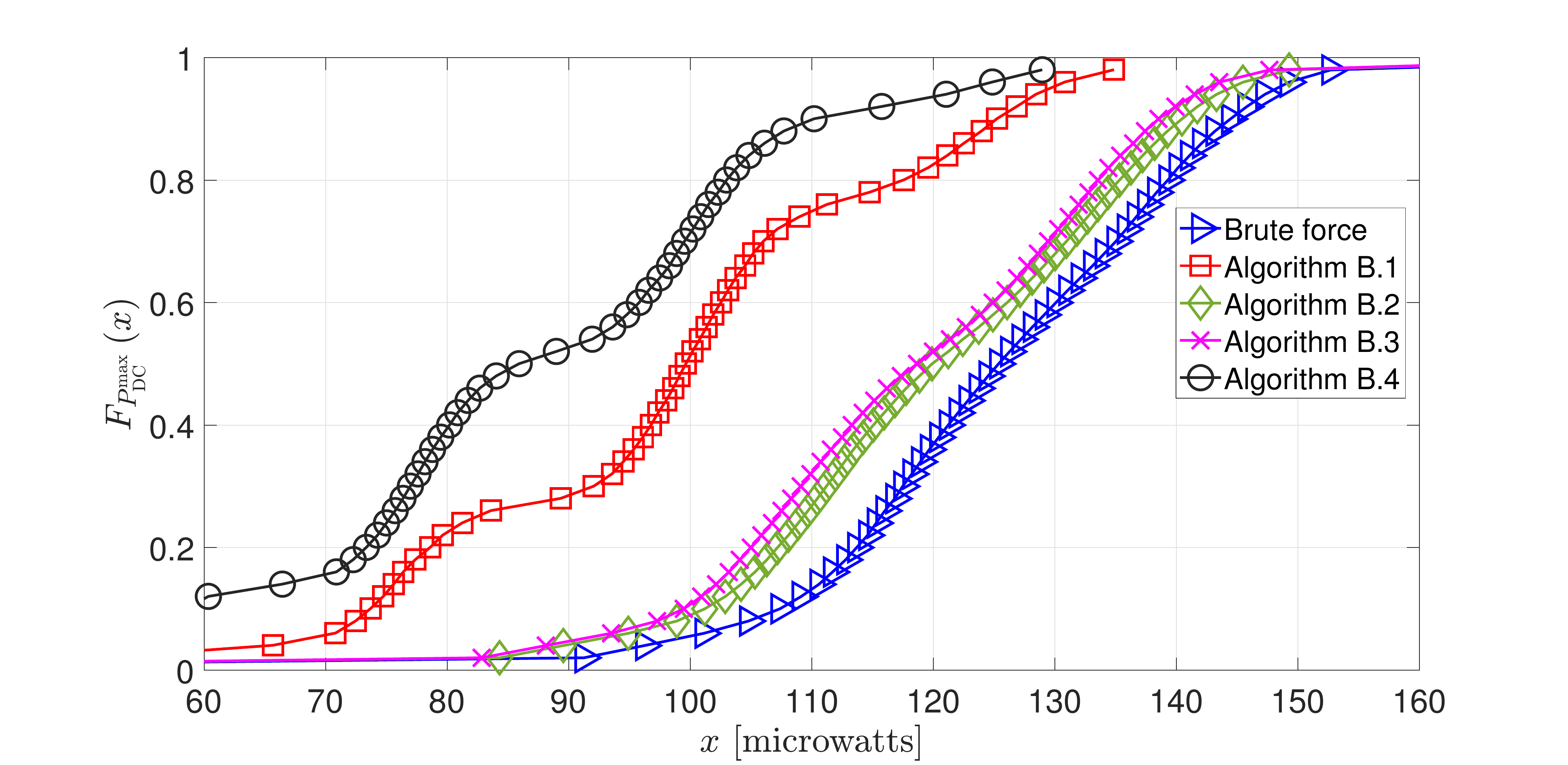}}\\
	(b) $\sigma_t^2=0.01$ ($K_1=100$).
	{\includegraphics[width=6in, height=3in]{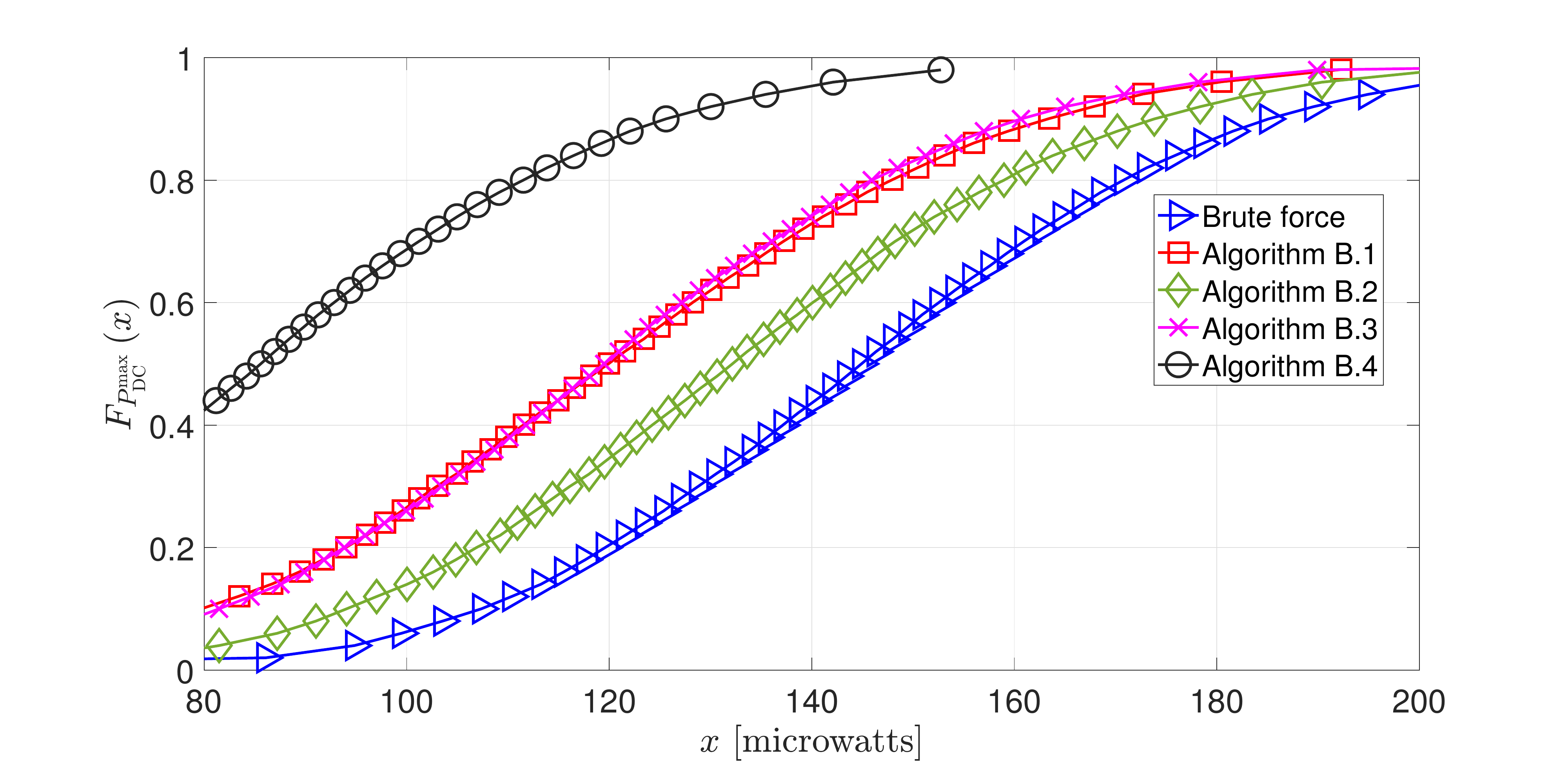}}\\
	(c) $\sigma_t^2=0.1$ ($K_1=10$).
	\caption{$F_{P_{\rm DC}^{\rm max}}\left(x\right)$ vs. $x$ for $M_x=5$, $M_y=2$ ($M_s=10$), and $\gamma_0=20$ dB.}
	%\hrulefill
	\label{Problem_B_CDF}	
\end{figure}

%\begin{table}[h]
	%\begin{small}
	%\caption{$ \bar P_{\rm{DC}}^{\rm{max}}$ for $M_x=4$, and $M_y=3$, $\gamma_0=18$ dB, and $K_1 \to \infty$.}
	%\begin{center}
	%\begin{tabular}{|c|c|}
		%\hline
		%Algorithm   & $ \bar P_{\rm{DC}}^{\rm{max}}$ [$\mu$W] \\ \hline
		%Brute force & 188.2                      \\ \hline
		%B.1         & 160                        \\ \hline
		%B.2         & 188.2                      \\ \hline
		%B.3         & 188.2                      \\ \hline
		%B.4         & 160                        \\ \hline
	%\end{tabular}
%\end{center}
%\label{Problem_2_Mean_Harvested_Power}
%\end{small}
%\end{table}

Regarding Problem B, as in Problem A, our first goal is to examine how close the performance of the proposed Algorithms B.1, B.2, B.3, and B.4 are to the corresponding one of the brute-force approach for varying channel conditions in the TX-RIS link ranging from free-space propagation to considerable scattering. Towards this end, in Fig.~\ref{Problem_B_CDF} we depict the cumulative density function of $ P_{\rm{DC}}(\mathcal{A}_
h^{\left(B\right)})$, denoted by $F_{P_{\rm DC}^{\rm max}}\left(x\right)$ and obtained by means of Monte Carlo simulations, for the brute-force approach and the proposed algorithms. As we observe, Algorithms B.2 and B.3 result in the same performance as the one of the brute-force approach for $\sigma_t^2=0$ and the performance gap increases with increasing $\sigma_t^2$, which validates our remark at the end of Section~\ref{Section_Problem_B_Solution}. In addition, among the proposed algorithms, Algorithm B.2 achieves the closest performance with respect to the corresponding one of the brute-force approach. The justification for this trend follows the same rational as the one for Algorithm A.1 in the case of Problem A since Algorithm B.2 also takes into account only the channel gains of the RIS-RX link.

\begin{figure}
	%\label{optimal_SNR_distance_vs_power_consumption}
	\centering
	{\includegraphics[width=6in, height=3in]{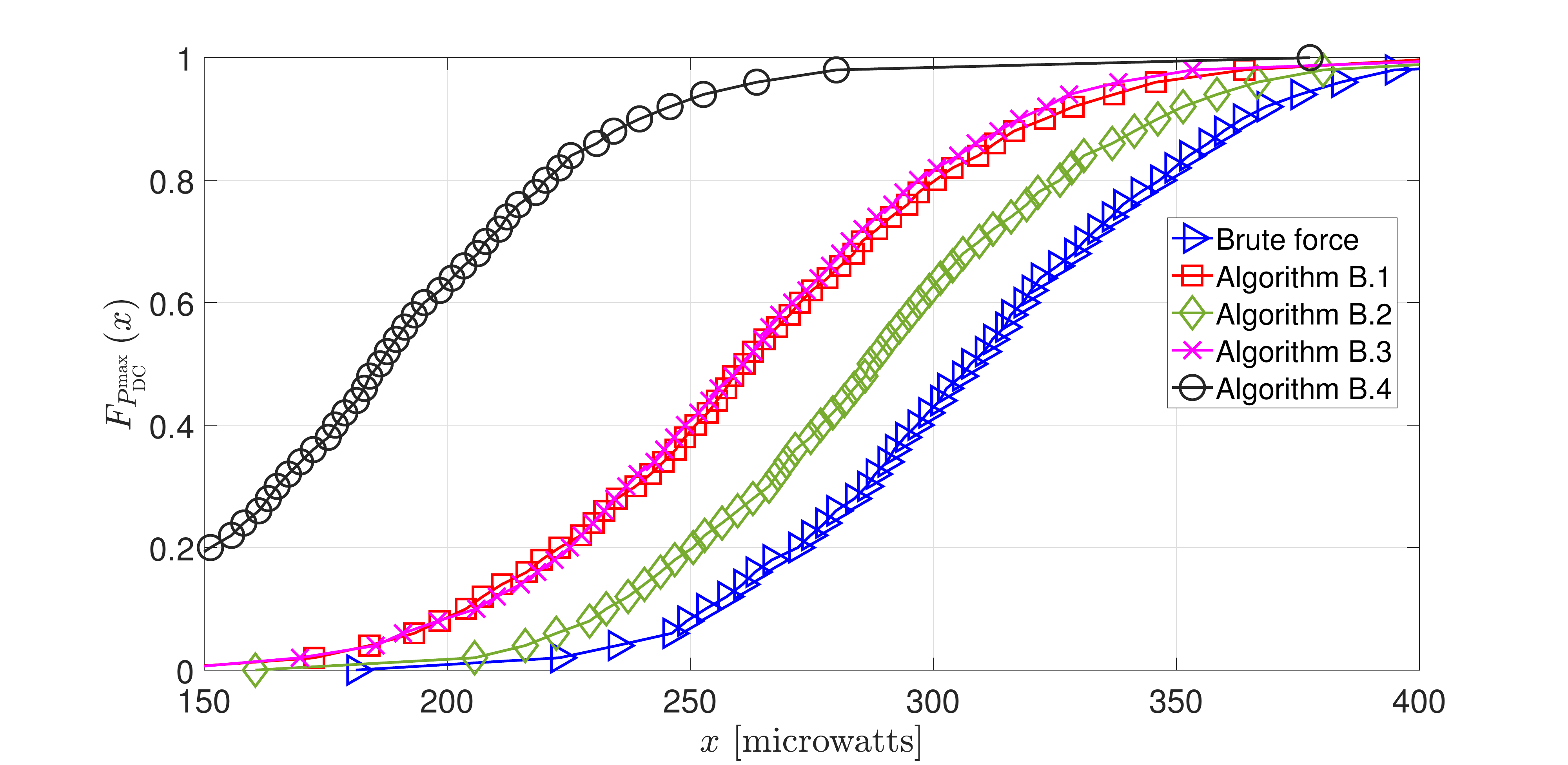}}
	\caption{$F_{P_{\rm DC}^{\rm max}}\left(x\right)$ vs. $x$ for $M_x=5$, $M_y=4$ ($M_s=20$), $\gamma_0=26$ dB, and $\sigma_t^2=0.1$ ($K_1=10$).}
	%\hrulefill
	\label{Problem_2_variance_0_1}	
\end{figure}

\begin{table}[h]
	\begin{small}
		\begin{center}
			\caption{Ratio of $\bar P_{\rm{DC}}(\mathcal{A}_
				h^{\left(B\right)})$ of the proposed algorithms over the one of brute force for and $\sigma_t^2=0.1$ ($K_1=10$). \label{Problem_2_Mean_Ratio_Average_Powers}}
			
			\begin{tabular}{|c|c|c|}
				\hline
				Algorithm & $M_x=5$, $M_y=2$ ($M_s=10$) & $M_x=5$, $M_y=4$, ($M_s=20$) \\ \hline
				B.1       & 82.9 \%                     & 85.2 \%                      \\ \hline
				B.2       & 91.5 \%                     & 93.5 \%                      \\ \hline
				B.3       & 82.8 \%                     & 84.6 \%                      \\ \hline
				B.4       & 59.8 \%                     & 60.6 \%                      \\ \hline
			\end{tabular}
		\end{center}
		
	\end{small}
\end{table}

Our second goal is to examine the relative performance gap of the proposed algorithms with respect to the one achieved by the brute-force approach as $M_s$ increases. Thus, in Fig.~\ref{Problem_2_variance_0_1} we depict $F_{P_{\rm DC}^{\rm max}}\left(x\right)$ for $M_x=5$, $M_y=4$ ($M_s=20$), and $\sigma_t^2=0.1$ ($K_1=10$). Comparing the plots of Fig.~\ref{Problem_2_variance_0_1} with the ones of Fig.~\ref{Problem_B_CDF}-(c) , where the latter refers to the $M_x=5$, $M_y=2$ ($M_s=10$) case, we observe that the performance gap of the algorithms with respect to brute force is almost constant, as in Problem A. To further substantiate this, in Table~\ref{Problem_2_Mean_Ratio_Average_Powers} we present the ratios of the average value of  $P_{\rm{DC}}(\mathcal{A}_
h^{\left(B\right)})$, denoted by $ \bar P_{\rm{DC}}(\mathcal{A}_
h^{\left(B\right)})$, of the proposed algorithms over the one of brute force for two values of $M_s$. As we observe, the ratios for the different $M_s$ remain almost constant, which indicates, as in the case of Problem A, that an increase of $M_s$ does not result in a performance loss of the proposed algorithms with respect to the one of brute force.

\subsection{Study-based discussion: User mobility tracking}

\begin{figure}
	%\label{Communication_through_an_RIS}
	\centering
	{\includegraphics[width=6in, height=3in]{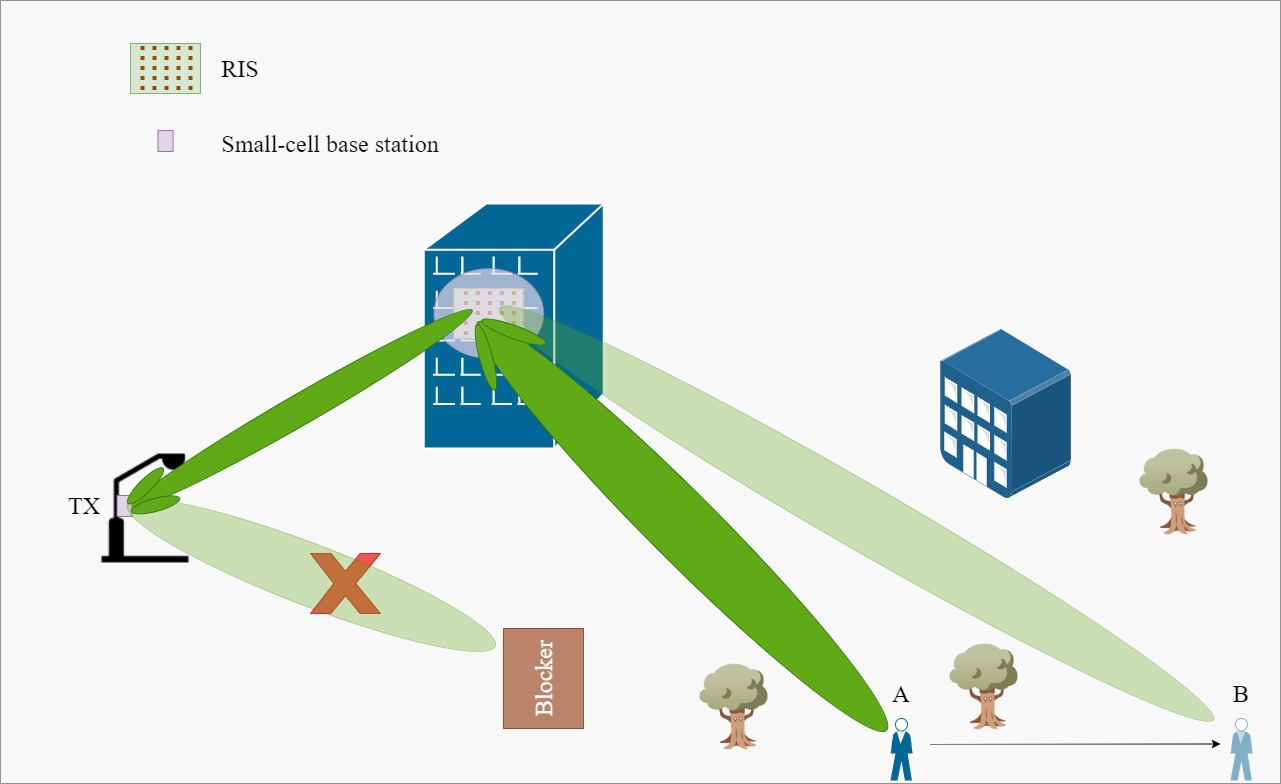}}
	\caption{User tracking through an RIS.}
	%\hrulefill
	\label{User_tracking}	
\end{figure}

One important question that arises based on the numerical results of Section~\ref{Numerical_results_subsection} is whether the corresponding $P_d^{\rm{avg}}$ in the order of few microwatts needed for autonomous RIS operation is attainable in practical scenarios. To assess this, let us consider a practical scenario where an RIS needs to be reconfigured in regular intervals. Such a scenario can indicatively be the tracking of a moving user, as it is depicted in Fig.~\ref{User_tracking}. In particular, a small-cell base station communicates with a user through an RIS. The user is moving along a straight line from point A to point B with a fixed velocity assumed to be equal to the average moving pace of human, which is around 1.4 m/s. In addition, we assume that the TX and RX antennas are located 3 at and 1.5 metres above the ground, respectively. Furthermore, we consider that the user is moving parallel to the RIS plane with a horizontal ground distance between the two planes equal to 17 m. Moreover, we assume the same horizontal ground distance between the TX and RIS. In addition, we assume a TX-RIS distance equal to 19 m. As far as the quality of the TX-RIS and RIS-RX links is concerned, for simplicity free-space propagation conditions are considered, i.e., $K_1, \; K_2\rightarrow\infty$. In addition, as a worst-case scenario, we assume $\alpha=1$. The rest of the parameters are the same as the ones considered in Table~\ref{Parameter_values}.

We consider the case where the origin is located at the point in the trajectory of the user that is closer to the RIS, and points A and B are located at $-40$ m and $40$ m, respectively. In addition, from a practical point of view, we rationally assume that the RIS is not instructed to be reconfigured continuously, but only when the end-to-end SNR drops below a specified threshold. A possible consideration for such a threshold could be the case when the SNR drops a certain amount of dB compared to the case of continuous tracking and reconfiguration. For illustration purposes, let us assume that such a threshold is 3 dB below the SNR value that would be achieved if continuous perfect tracking was employed. This represent the situation that the user moves beyond the half-power beamwidth of the current RIS configuration.

\begin{figure}
	%\label{Communication_through_an_RIS}
	\centering
	{\includegraphics[width=6in, height=3in]{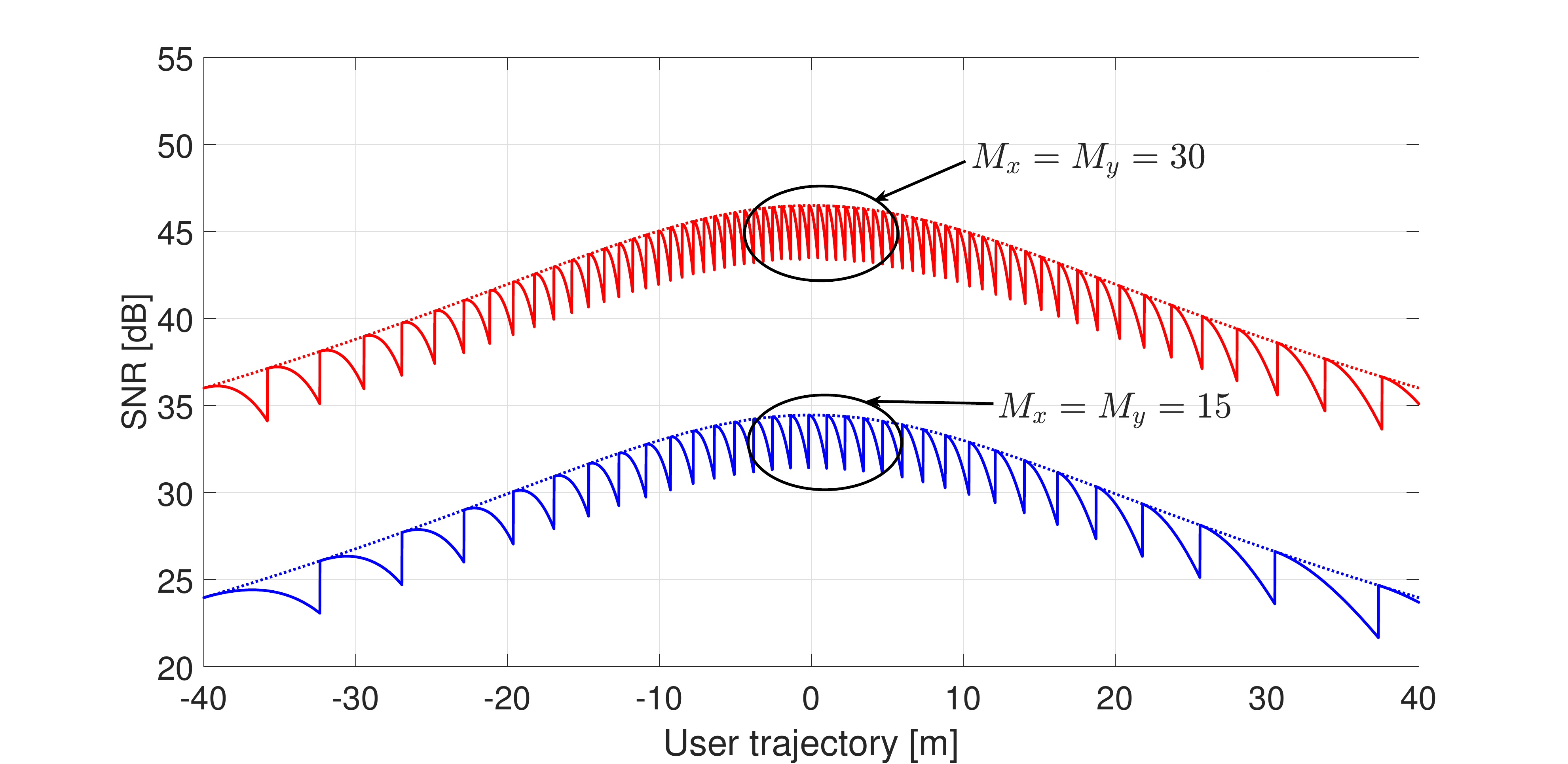}}
	\caption{SNR vs. user trajectory. Solid and dotted lines represent the practical periodic and ideal continuous reconfiguration, respectively.}
	%\hrulefill
	\label{Periodic_RIS_reconfiguration}	
\end{figure}

The resulting plots of periodic and continuous ideal reconfiguration are depicted in Fig.~\ref{Periodic_RIS_reconfiguration} for two values of $M_x, M_y$. As we observe, the closer to the RIS the user is, the more frequent the reconfiguration needs to be so that the SNR does not drop more than 3 dB compared to the SNR in the ideal case of continuous RIS reconfiguration. This was expected since, the closer the user is to the RIS, the smaller the main lobe footprint on the user plane is of the departing RIS beam. Hence, the SNR can drop 3 dB in smaller distances traveled by the user compared with the case that the user is further away from the RIS. More specifically, from Fig.~\ref{Periodic_RIS_reconfiguration}, we observe that when the user moves approximately in the $\left[-5\;\text{m}, 5\;\text{m}\right]$ range the RIS needs to be reconfigured every approximately 1.4 m (or 1 s) and 0.7 m or (or 0.5 s) in the $M_x=M_y=15$ and $M_x=M_y=30$ cases, respectively. On the other hand, when the user moves in the $\left[-40\;\text{m}, -30\;\text{m}\right]$ or $\left[30\;\text{m}, 40\;\text{m}\right]$ range the RIS needs to be reconfigured after 8 m (or 5.7 s) and after 4 m (or 2.85 s) in the $M_x=M_y=15$ and $M_x=M_y=30$ cases, respectively. The reconfiguration in the $M_x=M_y=30$ case needs to be faster than its $M_x=M_y=15$ counterpart due to the fact that the beamwidth of the departing RIS beam is smaller in the former case due to the higher number of UCs. This means that main lobe footprint on the user plane is also smaller in the specific case, which results in smaller moving distance from the user needed so that the SNR drops by 3 dB compared with the ideal case of continuous RIS reconfiguration.

Now, let us investigate how the reconfiguration duration of the RIS translates into power consumption values. To give a numerical example, by assuming a reconfiguration duration of several microseconds, for instance 100 $\upmu$s, which can be a typical value exhibited by RF-MEMS \cite{Paradigm_phase_shift}, it holds that $p_r=100 \; \upmu\text{s}/1\; \text{s}=10^{-4}$ and $p_r=100 \; \upmu \text{s}/0.5\; \text{s}=2\times 10^{-4}$ in the $M_x=M_y=15$ and $M_x=M_y=30$ cases, respectively, for the user trajectory closer to the RIS requiring the fastest reconfiguration. 

\begin{figure}
	%\label{Communication_through_an_RIS}
	\centering
	{\includegraphics[width=6in, height=3in]{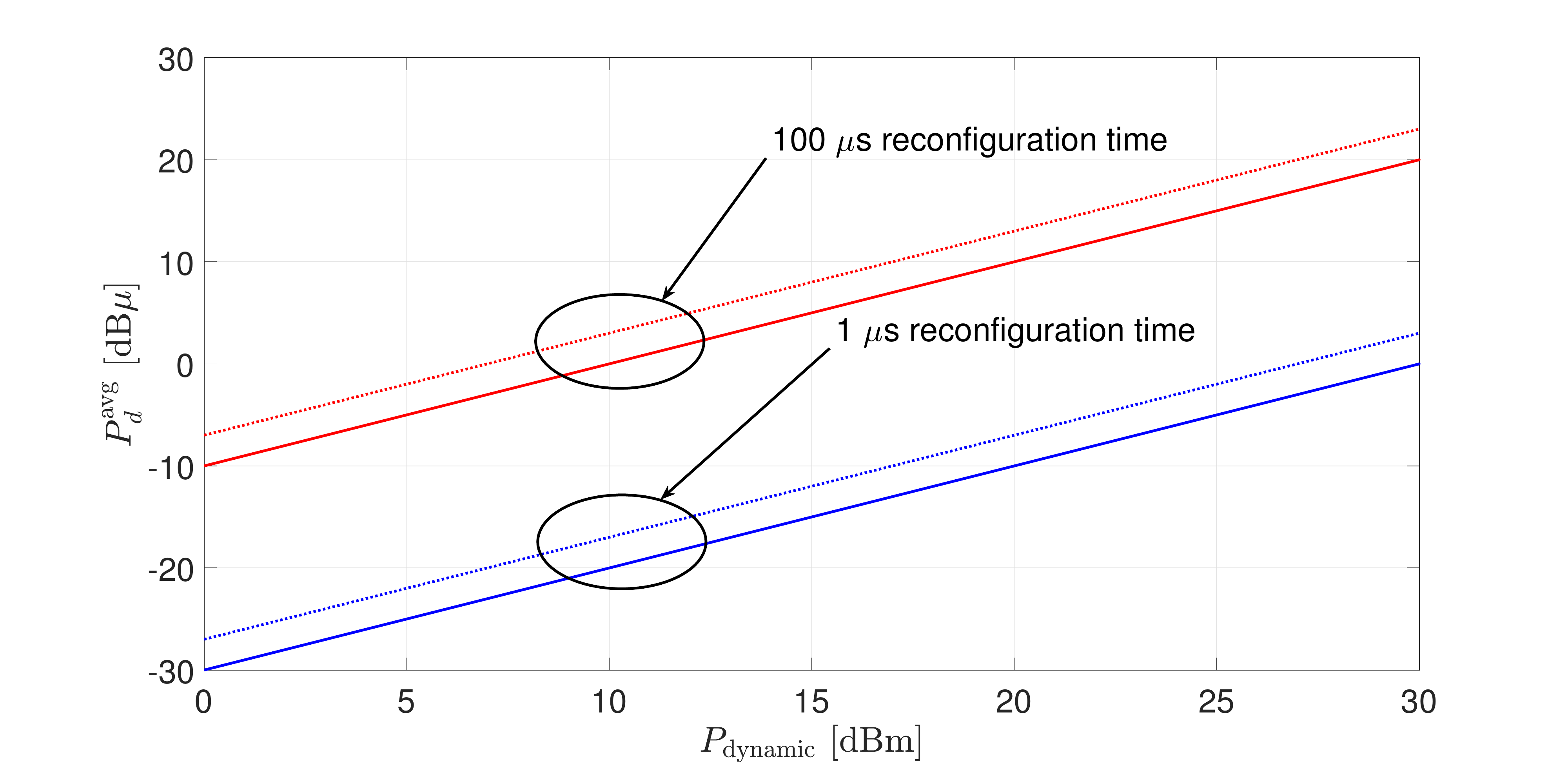}}
	\caption{$P_d^{\rm{avg}}$ vs. $P_{\rm{dynamic}}$. Solid and dotted lines represent the $M_x=M_y=15$ and $M_x=M_y=30$ cases, respectively.}
	%\hrulefill
	\label{Average_RIS_power_consumption}	
\end{figure}

The $P_d^{\rm{avg}}$ vs. $P_{\rm{dynamic}}$ plots for the $M_x=M_y=15$ and $M_x=M_y=30$ cases and for different reconfiguration durations are depicted at Fig.~\ref{Average_RIS_power_consumption}. As we observe, for the 1-$\upmu$s reconfiguration time, the maximum value of $P_d^{\rm{avg}}$ is 2 $\upmu$W for the $M_x=M_y=30$ case, which is below the 8 $\upmu$W value considered in the results of Section~\ref{Numerical_results_subsection}. This indicates that even if $P_{\rm{dynamic}}$ is relatively large, in the order of hundreds of milliwatts, a small reconfiguration time for the RIS electronics together with the requirement of not switching very fast can indeed keep $P_d^{\rm{avg}}$ in a level that can be secured by wireless energy harvesting from information signals. Such values of $P_{\rm{dynamic}}$ can be the case considering that the control chips need to also incorporate very low-power receiving units for receiving the wireless reconfiguration commands. We further add that a few RIS architectures have already been proposed where  electronic chips of ultra-low power consumption incorporate both the controlling unit consisting of a rudimentary central processing unit and the actuation network through the impedance-adjusting semiconductor components \cite[Fig. 3]{Liaskos_Design_SMM_2015}. Such chips are able to receive external wireless commands for reconfigurability and could also communicate with other RIS chips.

\section{Conclusions}

\label{Conclusions}

We have conducted this work to provide an answer to whether the widely argued RIS advantage of \emph{nearly-passive} operation compared to conventional active relaying can result in autonomous operation of the former by means of wireless energy harvesting from information signals. This is an important consideration because if the RIS technology requires a dedicated power supply, then the benefits over active relays in terms of deployment flexibility might be minuscule. To obtain a credible answer, we have identified the main RIS power-consuming modules and proposed a novel power-splitting architecture in which a subset of the RIS elements is allocated for wireless energy harvesting while the rest are used for beamsteering. Furthermore, we have formulated two optimization problems where the subset allocation is selected to either maximize the SNR under a constraint on the RIS power consumption or maximize the harvested power under a constraint on the end-to-end SNR. To solve these problems, we have provided low-complexity policies based on channel-gain ordering and proved that some of these deliver the optimal allocation under equal-gain conditions in the TX-RIS link. 

Numerical results have validated the close performance of some of the proposed policies with the one obtain by brute force. In addition, they have revealed the counter-intuitive outcome that the best performance for both problems is achieved by the algorithm that is based on channel-gain ordering of only the RIS-RX link. Finally, we have substantiated the potential of autonomous RIS operation by providing a case study of user mobility tracking. In particular, it was shown that autonomous RIS operation is feasible even under dynamic power consumption demands of several hundred milliwatts, which are realistic values based on current commercial hardware.

Overall, our work constitutes a first step toward the study of the autonomous RIS operation potential incorporating hardware aspects and the research needed regarding suitable scenarios, advancements in ultra-low power RIS electronic module design, and ways of wirelessly securing the required amounts of power. Future work will consider the performance comparison of the proposed power-splitting architecture with its time-splitting counterpart and explicit modeling of the channel estimation protocols.

\appendices

\section{Proof of Proposition~\ref{Proposition_1}}
\label{app:prop1}

Since $\left|h_{t_m}\right|^2=\beta$, for $m=1,2,...,M_s$, 
it follows that  $\sum_{i \in A_h}\left|h_{t_i}\right|^2=M_h \beta$. By solving \eqref{Problem_1_rewritten} with respect to $M_h$, we have
\begin{small}
	\begin{align}
		M_h=\left\lceil{\frac{-\left(\frac{1}{a}\right)\log\left(\frac{P_{\rm{max}}}{P_{\rm{RIS}}\left(1-\frac{1}{1+e^{ab}}\right)+\frac{P_{\rm{max}}}{1+e^{ab}}}-1\right)+b}{\eta_{\rm{RF}}
				P_t \beta
				%\left(\frac{\lambda}{4\pi}\right)^2
				%		\frac{P_{t}G_{t}G_{s}\left(\theta_{i}\right)}{d_t^2}
		}}\right\rceil.
	\end{align}
\end{small}
In addition, since every possible selection of $M_h$ energy harvesting UCs  out of the total $M_s$ UCs would result in the same harvested power due to the condition $\sum_{i \in A_h}\left|h_{t_i}\right|^2=M_h \beta$, the ones that are selected are the ones that associated with the $M_r$ lowest values of $\left|h_{r_m}\right|$ so that $\gamma\left(\mathcal{A}_r\right)$ is maximized. Based on this, the optimal solution of Problem A is given by Algorithm A.1 for $i_{\rm{stop}}^{\left(A\right)}$ given by \eqref{i_stop_Problem_A_free_space}. Furthermore, this optimal solution can also be reached by Algorithm A.2 due to fact that for $\left|h_{t_m}\right|=\sqrt{\beta}$, Algorithm A.2 degenerates to Algorithm A.1. This concludes the proof of Proposition~\ref{Proposition_1}.

\bibliographystyle{IEEEtran}
\footnotesize{
	\bibliography{IEEEabrv,references}
}

\end{document}